\documentclass[aps,pra,a4paper,twocolumn,showpacs,eqsecnum]{revtex4}
\usepackage{graphicx}

\begin{document}
\title{Quantum information approach to the Ising model: Entanglement in chains of qubits}
\author{Peter \v{S}telmachovi\v{c}${}^1$ and Vladim\'{\i}r Bu\v{z}ek${}^{1,2}$}
\affiliation{${}^1$ Institute of Physics, Slovak Academy of Sciences,
D\'{u}bravsk\'{a} cesta 9, 845 11 Bratislava, Slovakia \\
${}^2$ Faculty of Informatics, Masaryk University, Botanick\'{a} 68a,
602 00 Brno, Czech Republic}

\begin{abstract}
Simple physical interactions between spin-1/2 particles may result
in quantum states that exhibit exotic correlations that are
difficult to find if one  simply explores state spaces of
multi-partite systems. In particular, we present a detailed
investigation of the well known Ising model of a chain (ring) of
spin-1/2 particles (qubits) in a transverse magnetic field. We
present explicit expressions for eigenstates of the model
Hamiltonian for arbitrary number of spin-1/2 particles in the
chain in the standard (computer) basis and we investigate quantum
entanglement between individual qubits. We analyse bi-partite as
well as multi-partite entanglement in the ground state of the
model. In particular, we show that bi-partite entanglement between
pairs of qubits of the Ising chain (measured in term of a
concurrence) as a function of the parameter $\lambda$ has a
maximum around the point $\lambda=1$ and it monotonically
decreases for large values of $\lambda$. We prove that in the
limit $\lambda\rightarrow\infty$ this state is locally unitary
equivalent to an $N$-partite Greenberger-Horn-Zeilinger state. We
also analyse a very specific eigenstate of the Ising Hamiltonian
with a zero eigenenergy (we denote this eigenstate as the
$X$-state). This $X$-state exhibits the ``eXtreme'' entanglement
in a sense that an arbitrary subset $A$ of $k\leq n$ qubits in the
Ising chain composed of $N=2n+1$ qubits is maximally entangled
with the remaining qubits (set $B$) in the chain. In addition we
prove that by performing local operation just on the subset $B$
one can transform the $X$-state into a direct product of $k$
singlets shared by the parties $A$ and $B$. This property of the
$X$-state can be utilised for new secure multi-partite
communication protocols.
\end{abstract}

\maketitle

\section{Introduction}

  Those multi-partite quantum systems which are fundamental objects of
  statistical and solid state physics, have been found interesting also
  from a perspective of quantum information processing.
These systems often exhibit multi-partite entanglement that can be used either
for quantum information processing or quantum communication.
Amongst such systems  a distinguished role is played by
  exactly solvable models, such as the Ising model describing a chain of interacting
  spin-1/2 particles in an external magnetic field.
Eigenstates of the corresponding model Hamiltonian can be studied
from a perspective of quantum information theory with a good physical
motivation:
Any quantum computer is a
physical device composed of elementary units, qubits, described
by a certain Hamiltonian. Consequently, perfect
knowledge of the Hamiltonians and their eigenvectors are vital.
An important condition the physical system
has to fulfil is the possibility of preparation of an
{\em a priori} known initial state. The easiest way to realize this task is to simply let the system
evolve into its ground state. Thus the knowledge of the
entanglement properties of the ground state or more practically
thermal states are necessary. This  has been
followed by many authors. In particular,
various versions of the
  Heisenberg model (XX, XY, XYZ) have been investigated. Many of these studies concern
   numerical and
 analytical investigations primarily focused on the behaviour of
  \emph{bipartite} entanglement of small  number
  of qubits in ground and thermal states, e.g.
  Refs.~\cite{ArnesenBV01,WangFS01,Wang01,Wang01a,GunlyckeKVB01,XuZZN02,XiHCY02,XiHCY02,Wang02,Wang02a,FuSW02,ZhouSGL03,fusowa}.
  The notion of ``thermal entanglement'', i.e. the entanglement of
  thermal states is introduced, and its properties including threshold
  temperatures and magnetic field dependence are studied.

  Symmetry properties of multi-partite systems have been used to calculate
  entanglement among their constituents. In Ref.~\cite{Schliemann03} thermal
  equilibrium states of isotropic two-spin systems are analysed
  exploiting SU(2) invariance. The results are related to isotropic
  Heisenberg models. In Ref.~\cite{GlaserBF03}, analytical expressions
  for certain entanglement measures are derived using general
  symmetries of the quantum spin system. Then they are used for the
  XXZ model in order to calculate concurrence and the critical
  temperature for disentanglement for finite systems with up to six
  qubits. It should be noted, that they use the 3-tangle to analyse
  some multi-partite entanglement aspects of the system, and discuss
  entanglement sharing in detail.

   In Ref.~\cite{BriegelR01} the authors pointed out, that in a
    finite chain of qubits, the time evolution generated by the Ising
    Hamiltonian produces ``entanglement oscillations'', which lead to
    the presence of GHZ and W type entangled states. A generalisation
    to 2D and 3D models is also outlined. Discussions of multi-partite
    entanglement also appear  in
    Refs.~\cite{WangFS01,Wang01,Wang02a}. In Ref~\cite{Yeo03} quantum
    teleportation is utilised as a tool to reveal the importance of
    multi-qubit entanglement in a 3 qubit Heisenberg-XX chain.

  A central question in the problem of entanglement of more than two
  systems is that of bounds on
  entanglement. Three or more quantum systems cannot be arbitrarily
  entangled in the similar way as they cannot be arbitrarily
  classically correlated~\cite{CoffmanKW00}. The state with an {\em a priori} specified entanglement
    properties may not exits at all~\cite{PleschB03,PleschB03a} and
  therefore the search for a state with
    given, in a sense optimal entanglement properties, is in general a
  hard problem. In Refs.~\cite{Wootters2,OConnorW01} the
  authors have solved  such particular
  problems by minimising the energy of a Hamiltonian.
  That is, the sought state with a given pairwise entanglement
  is the ground state of a Hamiltonian with a very clear physical
  interpretation. Since such
  states may be useful for quantum information
    processing it is desirable to known the concrete form of
  the states that are either
    optimal, or obey certain bounds. It
    should be pointed out that the
    problem of finding entangled webs with given properties has been
  extensively addressed in Refs.~\cite{KoashiBI00,PleschB03} but
  without any reference to systems described with Hamiltonians.
   Thus it is
  interesting to see how
  this issue can be approached in other exactly solvable models.

Another interesting  issue concerning such models of  many-body systems
is the  collective behaviour of these systems under certain conditions
known as the critical phenomena.
Let us point out that these phenomena has already been studied
extensively.
On the other hand it has been pointed out only
recently~\cite{Osterloch,OsborneN02,VidalLRK03,GhoshRAC03,Vidal3} that
  entanglement is the quantity that {\it may}  play a crucial role in
  the description and understanding of critical phenomena.
 The central concept of the theory of critical phenomena is the
 universality - the critical exponents characterising divergences near
 critical points are the same for all systems belonging to the
 same universality class.
For a special class of one-dimensional magnetic systems
  it has  been shown in Ref.~\cite{Osterloch} that the bipartite
  entanglement shows scaling behaviour near the transition point.
  One   should also expect, that precursors of the critical behaviour may
  appear even in non-critical systems.

In this paper
we present a detailed
investigation of the well known Ising model of a chain (ring) of spin-1/2
particles (qubits) in a transverse magnetic field (Section II). We present
explicit expressions for eigenstates of the model Hamiltonian for
arbitrary number of spin-1/2 particles in the chain in the
standard (computer) basis and we investigate quantum entanglement
between individual qubits (Sections III and IV). We analyse bi-partite as well as
multi-partite entanglement in the ground state of the model. In
particular, we show that bi-partite entanglement between pairs of
qubits of the Ising chain
(measured in term of a concurrence defined in Section II) as a function of the parameter
$\lambda$ has a maximum around the point $\lambda=1$. In addition, it
monotonically decreases for large values of $\lambda$. We prove
that in the limit $\lambda\rightarrow\infty$ this state is locally
unitary equivalent to an $N$-partite Greenberger-Horn-Zeilinger
state (Section IV).
We also analyse a very specific eigenstate of the Ising Hamiltonian with a
zero eigenenergy (we denote this eigenstate as the $X$-state). This $X$-state
exhibits eXtreme entanglement in a sense that a arbitrary subset $A$ of $k\leq n$ qubits in
the Ising chain composed of $N=2n+1$ qubits is maximally entangled with the remaining qubits (set $B$)
in the chain. In addition we prove that by performing local operation just on the subset $B$ one can
transform the $X$ state into a direct product of $k$ singlets shared by the parties $A$ and $B$. This
property of the $X$ state can be utilised for new secure multi-partite communication protocols.
 Technical details of some of  our calculations are presented in appendices.

\section{Setting-up the scene}

\subsection{The Ising model}

We  consider a model of a linear chain of spin one-half particles
forming a circle, placed in a magnetic field where only the
z-component of the field is non-zero. Since we are interested in the
spin degrees of freedom only, the Hamiltonian of the system is given
by
\begin{equation}
\label{6.1.1}
 {H}_N = - C_I \sum_{n=1}^{N} \sigma_{n}^{x} \otimes \sigma_{n+1}^{x} + B \sum_{n=1}^{N} \sigma_{n}^{z} \; ,
\end{equation}
where $\sigma_n^{\alpha}, \alpha=x,y,z$ are well known Pauli
operators. The first term in the Hamiltonian is the interaction term
with coupling constant $C_I $ and the second term corresponds to a
free Hamiltonian. The lower index $ n $ labels the position of a spin
in the chain and $ N $ is the overall number of particles. The cyclic
boundary conditions
\begin{equation}
\label{6.1.2}
\hspace{1cm} \sigma_{N+1}^{\alpha} = \sigma_{1}^{\alpha}\, ; \hspace{2cm} \alpha=x, y, z
\end{equation}
ensure that the chain forms a circle. The form of the interaction is
chosen such that each particle interacts only with its two nearest
neighbours.

The Hamiltonian in
Eq.~(\ref{6.1.1}) can be rewritten  into a form which
is more convenient for numerical calculations:

\begin{equation}
\label{6.1.3}
 {H}_N = E \left \{ - \lambda \sum_{n=1}^{N} \sigma_{n}^{x}
\otimes \sigma_{n+1}^{x} + \sum_{n=1}^{N} \sigma_{n}^{z} \right \} \; ,
\end{equation}
where $E=B$ and $\lambda=C_I/{B}$ is a dimensionless parameter. Now,
instead of two parameters in energy units ($B$ and $C_I$) we have one
dimensionless parameter $\lambda$ and one parameter in energy units,
$E$, that can be neglected in our further calculations except for the
investigation of entanglement in Gibbs states (see Section
\ref{s6.2.3}).

\subsection{Measures of entanglement}
\label{sect:entang}

In this paper we will use three different measures --
the concurrence, the tangle and a measure of an intrinsic three-partite entanglement.

The concurrence \cite{Wootters} is a measure of the bipartite entanglement
between two qubits.  Let $\rho_{AB}$ be the joint
density matrix of the system consisting of qubits
  $A$ and $B$. The matrix $\rho_{AB} \tilde{\rho}_{AB}$ has four non-negative
eigenvalues $\{ \lambda_1, \lambda_2, \lambda_3, \lambda_4 \}$ that are written in a
descending order (i.e. $\{ \lambda_1 \geq \lambda_2\geq \lambda_3\geq \lambda_4 \}$.
The matrix
$\tilde{\rho}_{AB}$ is a spin-flipped version of
 density matrix $\rho_{AB}$, i.e. $\tilde{\rho}_{AB}=\sigma_y\otimes\sigma_y \rho_{AB}^* \sigma_y\otimes\sigma_y$. The
concurrence is given by the relation
\begin{equation}
\label{3.4.2.1}
C_{AB} = {\rm max} \left\{\left[ \sqrt{\lambda_1}  - \sum_{i=2}^4 \sqrt{\lambda_i}\right]
; 0\right\} \; .
\end{equation}
Let us point out that the state $\rho_{AB}$ is arbitrary, that is, the concurrence is
a valid measure of entanglement for two-qubit mixed states as well. The minimal value of the concurrence
is zero (in this case two qubit states are prepared in a separable state), while for any $C_{AB} > 0$ two qubits
are entangled. The maximal value of $C_{AB}=1$ corresponds to maximally entangled states that are locally unitary
equivalent to Bell states.
It has been shown earlier that
the concurrence is directly related to the entanglement of formation \cite{Wootters}.

On the other hand the tangle  has been originally defined for pure states only. (There
is an extension to mixed states but the extension is not
computationally feasible except for the case of two qubits when the tangle
is equal to the square of the concurrence). Consider a multi-partite system where one
of the subsystems, denoted $A$, is a qubit. The tangle $C_{A\bar{A}}$ between
a subsystem $A$ and the rest of the system, denoted as $\bar{A}$
reads
\begin{equation}
\label{3.4.1.2}
C_{A\bar{A}} = 4 \: {\rm Det} \rho_A = 2 (1 - {\rm Tr} \rho_A^2 ) \; ,
\end{equation}
where $\rho_A$ is the reduced density operator describing a state of the subsystem $A$.

Finally, the intrinsic three-qubit entanglement is defined for pure states only.
 Consider a system composed of three qubits $A$, $B$ and $C$ and
let the system of three qubits be in a pure state. With the help of
the tangle and concurrence introduced above we can define pure
three-partite entanglement \begin{equation}
\label{3.4.3.3}
C_{ABC} \equiv  C_{A\bar{A}} - C^2_{AB} - C^2_{AC} \; .
\end{equation}
In Ref.~\cite{CoffmanKW00} it has been proven that the definition
(\ref{3.4.3.3}) of intrinsic three-qubit entanglement is independent of permutations of particles and
shares all properties that a proper measure of entanglement has to fulfil.

Finally, let us mention that measures of entanglement are not unique
and different measures might result in different ordering of states. For the case of bipartite systems
$AB$ prepared in a pure state the measure of entanglement is in fact any suitable
function  of the eigenvalues of the reduced density matrix of either of the two
subsystems $\rho_A$ or $\rho_B$  Ref.~\cite{Horodecki}. For example the well known Von Neumann
entropy
\begin{equation}
\label{entropy}
S(\rho_A) = -  {\rm Tr} \rho_A \log \rho_A \; ,
\end{equation}
defines a  bipartite measure of entanglement usable for arbitrarily dimensional
systems.

\section{Three spin-1/2 particles}
\label{s6.2}
In order to understand entanglement properties of the Ising chain under consideration
it makes sense to start with a relatively simple example of three spin-1/2 particles (qubits).
Even this simple system exhibits interesting properties and their understanding
will guide us in general case of an arbitrary number of qubits.

The Ising Hamiltonian with three spin-1/2 particles can be directly diagonalized and energy
levels easily calculated. In what follows, we will call spin-1/2 particles as qubits
since the Hilbert space ${\cal H}$ of a spin one-half particle is
two-dimensional. Let us note that the simplest example is the case of
two qubits, that is $N=2$.  However, this trivial example has already
been investigated in Refs. \cite{ArnesenBV01} and
\cite{GunlyckeKVB01}.  The case of three qubits is also interesting on account of
the fact that, besides intrinsic
bipartite entanglement, three qubits can also share
three-partite entanglement. In
the case of three qubits being in a pure state this intrinsic three-qubit entanglement can
be easily
calculated with the help of Eq.~(\ref{3.4.3.3}). Finally, even such a
simple example nicely illuminates main results concerning
multi-partite entanglement where most results can be
generalised to the case with an arbitrary number $N$ of
qubits (spin-1/2 particles) in the chain. \vskip 0.3cm

\noindent The Hamiltonian Eq.~(\ref{6.1.3}) of the Ising model with
only three qubits  in the chain reads
\begin{equation}
\label{6.2.1}
H_3 = - \lambda \sigma_1^x \otimes \sigma_2^x - \lambda
\sigma_2^x \otimes \sigma_3^x - \lambda \sigma_1^x
\otimes \sigma_3^x + \sigma_1^z + \sigma_2^z + \sigma_3^z \; .
\end{equation}
Note that the Hamiltonian $H_3$ is permutationally invariant, unlike
the Hamiltonians $H_N$ for $N > 3$
qubits. All Hamiltonians $H_N$
are obviously translationally invariant. What is not so obvious is the fact
that the Hamiltonian is invariant under the inversion
of the order of particles. The particles in the chain are labelled
with $n=1, \ldots N$. Now, if we relabel them as
$n \rightarrow N - n + 1$ the Hamiltonian remains unchanged and thus
is invariant under the inversion of the order. For
the case of $N=3$ the two transformations, translation and
inversion of the order, together with an arbitrary
combination of the two yield in fact all possible
permutations of the particles in the chain. Thus it
follows that the Hamiltonian $H_3$ is permutationally invariant.

The knowledge of
  the symmetry of the Hamiltonian is utmost important, as it plays a
  crucial role in the process of finding its eigenvalues and
  eigenvectors. It is well known
that for every symmetry $S$ there exists a unitary or an anti-unitary
operator  $T_S$ such that the corresponding
Hamiltonian commutes with $T_S$
\begin{equation}  [T_S, H] = 0 \;.
\end{equation}
As a result of this commutation relation
the two operators $H$ and $T_S$ have common set of
eigenvectors. It means, that there is one set of vectors (basis of the
corresponding Hilbert space) which are eigenvectors of $H$ as well as
the operator $T_S$. Moreover, any non-degenerate eigenstate of $H$
has to be invariant under the action of the operator $T_S$. On the
other hand, any eigenstate which is not invariant under the action of
the operator $T_S$ is degenerate. In what follows the knowledge of
symmetries of $H_N$ will help us to find some particularly interesting
states of the spin-chain under consideration.

\subsection{Spectrum of the Hamiltonian}
\label{s6.2.1}

The spectrum of the Hamiltonian $H_3$ can be easily calculated directly by
diagonalising the Hamiltonian $H_3$. The Hilbert space
${\cal H}_2 \otimes {\cal H}_2 \otimes {\cal H}_2$ of three qubits is
eight dimensional and the Hamiltonian $H_3$ has eight eigenvalues (see
Appendix~\ref{p3vhvv}), shown in Fig.~\ref{spectrum-3} as functions~\footnote{
The energy levels $E_j, j=1, \ldots 8$  are expressed in energy units defined by the parameter
$E$ of the Hamiltonian.
In what follows  this parameter is dropped
and the energy levels $E_j$ are taken to be real
functions of the parameter $\lambda$.}
 of the coupling
constant $\lambda$.
\begin{figure}
\includegraphics[width=7cm]{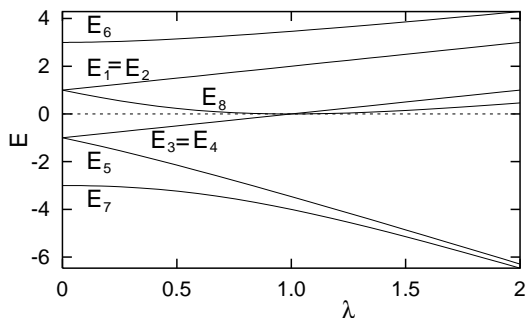}
\caption{The spectrum of the Hamiltonian $H_3$. We present the dependence
of eigenenergies as a function of the coupling parameter
$\lambda$. The energy levels $E_1=E_2$ and
$E_3=E_4$ are degenerate. The other four energy levels are non-degenerate.
The ground state corresponding to the state with the lowest energy
in our notation is represented by  the seventh level
$E_7$ for all values of the parameter $\lambda$.}
\label{spectrum-3}
\end{figure}
Two of them are double degenerated, while the remaining
four are not, apart from several isolated values of the parameter
$\lambda$. The ground state of the system for any finite value of
$\lambda$ is non-degenerate, and in our notation it is the
seventh state $|e_7 \rangle$. When the parameter $\lambda$ is
infinite, which corresponds to the  zero value of
the external magnetic field $B$, the Hamiltonian $H_3$ has only one
free parameter, and can be expressed as
\begin{eqnarray*}
H_3 (\lambda = \infty) = - C_I \left ( \sigma_1^x \otimes \sigma_2^x + \sigma_2^x \otimes \sigma_3^x + \sigma_1^x \otimes \sigma_3^x \right ) .
\end{eqnarray*}
The two lowest states (denoted as $| g_1 \rangle$ and $|g_2 \rangle$)
in the energy spectrum become degenerate in this case. These states read
\begin{eqnarray*}
|g_1 \rangle = \frac{1}{2} (|000 \rangle + |110 \rangle + |011 \rangle + |101 \rangle ) \; , \\
|g_2 \rangle = \frac{1}{2} (|111 \rangle + |100 \rangle + |010 \rangle + |001 \rangle ) \; . \\
\end{eqnarray*}
Here the first one is the limit of the state $|e_7
\rangle$ when $\lambda$ tends to infinity and the second one is  the
limit  of the state $|e_5 \rangle $. We know that any linear
combination of the two vectors is an eigenvector
with the same energy and, consequently, can be identified as a ground
state. However, there is one exceptional linear combination. If we
demand the ground state of the system with the parameter
$\lambda=\infty $ to be the limit of the ground state when
$\lambda=\infty $, then the
appropriate choice for the ground state is $|g_1 \rangle$.

As we will see in Section~\ref{s6.3.1}, the point
$\lambda=1$ turns out to be rather interesting. There is a particular
eigenstate of the Hamiltonian which has quite interesting behaviour of entanglement around
$\lambda=1$. However, for $N$ large it is rather difficult to identify
this specific state among $2^N$
eigenstates of the Hamiltonian $H_N$. Having calculated the spectrum, the state can be
easily identified with the help of the level crossing at the point
$\lambda=1$. The special state with the remarkable properties
is in general non-degenerate but at the point $\lambda=1$ becomes
degenerate
\begin{eqnarray*}
E_8 (\lambda=1) = E_{3,4} (\lambda=1) = 0 \; ,
\end{eqnarray*}
and crosses the degenerate levels $E_{3}$ and $E_4$. What is important
is the fact that this type of level crossing is independent of $N$ (we
might say universal),  there is the same type of level
crossing for $N$ being an arbitrary odd number.

\subsection{Entanglement properties}
\label{s6.2.2}

Our main goal is to analyse the
entanglement properties of the model.  Let us begin with
the Ising chain of three qubits in the ground state and examine entanglement as a function of the parameter
$\lambda$. We will use the three different measures of entanglement: The concurrence, the tangle and
  a measure of the intrinsic three-partite entanglement, as introduced in
  Section~\ref{sect:entang}. An important aspect is the comparison of
  specifically bipartite, and multi-partite entanglement.

\begin{figure}
\includegraphics[width=7cm]{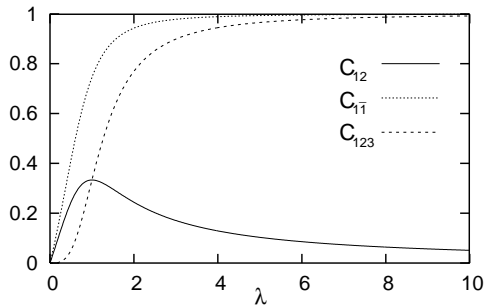}
\caption{The entanglement
in the ground state of the Ising chain with three qubits $N=3$. In the
figure we present
three different types of the entanglement as a function of the parameter
$\lambda$: the bipartite entanglement between the first and the second qubit
$C_{12}$ (solid line); the bipartite entanglement between the first  qubit
and the
remaining two qubits $C_{1\bar{1}}$ (dotted line) and the intrinsic three-partite
entanglement $C_{123}$ (dashed line).}
\label{obr-n3zs}
\end{figure}

The bipartite entanglement between individual qubits, the entanglement between
a qubit and the rest of a system, and an intrinsic
three-partite entanglement for the ground state are shown in
Fig.~\ref{obr-n3zs}.
They are quantified by concurrence,
  tangle, and the intrinsic three-partite entanglement of
  Eq.~(\ref{3.4.3.3}), respectively. Due to the fact that any non-degenerate
state shares all symmetries of the corresponding Hamiltonian, the
entanglement of the ground state between an arbitrary pair of qubits
has the same dependence on the parameter $\lambda$ and it holds that
\begin{equation}
C_{12} (\lambda) = C_{13} (\lambda) = C_{23} (\lambda) \; ,
\end{equation}
Moreover, the same holds for bipartite entanglement between a given
qubit and the rest of the system so that
\begin{equation}
C_{1 \bar{1}} (\lambda) = C_{2 \bar{2}} (\lambda) =
C_{3 \bar{3}} (\lambda) \; ,
\end{equation}
where $\bar{X}$ denotes a system of two qubits with the qubit on the $X$-th position
omitted and $C_{X \bar{X}}$ is the entanglement
shared between the qubit on the $X$-th position and the rest of the
Ising chain (remaining two qubits).

The solid line in the picture Fig.~\ref{obr-n3zs} shows bipartite entanglement between
an arbitrary pair of qubits. For $\lambda=0$ the concurrence, i.e. the pairwise entanglement is
zero. As $\lambda= C_I/B$ it means that the case $\lambda=0$
corresponds to the absence of the  interaction, $C_I=0$. Consequently, the ground
state of the system is such that all spins are aligned along the same
direction, the direction of the magnetic field, and are not entangled.
When we turn on the interaction, the constant $C_I$ is no longer zero
and the spins become entangled. As we increase the value of the
interaction constant $C_I$ (or equivalently decrease the value of the
magnetic field $B$ so that the ratio $C_I/B$ increases) the qubits
become more and more entangled. This holds up to the value of $\lambda=1$
where the two qubit entanglement reaches its maximum. Further
increase of the parameter $C_I$ (or decrease of $B$) causes
degradation of the entanglement and in the limit $\lambda$ goes to
infinity the entanglement is zero. As we have chosen the ground state
for $\lambda=\infty$ to be the limit of the ground state when
$\lambda$ tends to infinity, the concurrence at the point
$\lambda=\infty$ is zero. It means that when the magnetic field is
zero, the ground state of the system is such that all pair
concurrencies are zero and there is no entanglement in any pair of
qubits.

The entanglement of a given qubit $X$ and the rest
of the system $\bar{X}$ expressed in terms of the
tangle is, on contrary, a non-decreasing function of $\lambda$.
At the point $\lambda=0$ the entanglement is zero on account of the
same reason as the entanglement between an arbitrary pair of qubits.
When the parameter $\lambda$ is non-zero (i.e. the interaction
constant $C_I$ is non-zero) the qubits are entangled. That is,
any chosen individual qubit is entangled with the rest
of the system.  The
stronger the interaction (the larger the value of the parameter
$\lambda$), the stronger the qubits are entangled with the system. In
the limit $\lambda\to\infty$ (infinitely strong interaction) the
qubits become maximally entangled and the tangle, measuring the amount
of entanglement between a given qubit and remaining two qubits,
reaches its maximum value. For the case of our specific choice of the
ground state for $\lambda=\infty$ the tangle is maximal and equals to unity.

As we have already pointed out the reason why we have described the
case of three qubits in such detail is that the entanglement behaves
in the same manner for an arbitrarily large $N$. But
the case of three qubits is special for a different reason too.
In the case of just three qubits being in a pure state we are able to
calculate the intrinsic three-partite entanglement using Eq.~(\ref{3.4.3.3}).
The dashed line in the Fig.~\ref{obr-n3zs} shows the dependence of the three-qubit
entanglement on the parameter $\lambda$. We can see that the
dependence of the intrinsic three-partite entanglement on the parameter
$\lambda$ is very similar to the dependence on the same parameter of
the entanglement between a given qubit and remaining two qubits
(Fig.~\ref{obr-n3zs}). It seems that for a strong interaction the
three-partite entanglement is the largest
contribution provided we express the entanglement between a single
qubit and remaining two qubits (rest of the system) as a sum of two
and three-qubit entanglement (see Eq.~(\ref{3.4.3.3}) and comments
therein). This result suggests the following physical
  picture:\emph{
  ``When the system of interacting spin-1/2 particles is in the ground state then
    the interaction causes  entanglement of qubits such that
  each qubit is entangled with the rest of the system.  For the system of $N$ spin-1/2 particles the
  $N$-partite entanglement will be dominant when
  the interaction between the particles is very strong compared to the
  magnitude of the magnetic field.''}
This conjecture, proven to be valid in the case of
three qubits, will be further examined in following sections where the
general case of a chain with an arbitrary number of qubit will be
analysed.

\subsection{Entanglement in Gibbs ensembles}
\label{s6.2.3}

In this subsection we will continue to investigate the 3-qubit Ising model. We will analyse the
  entanglement properties of thermal states of three qubits interacting according to the Ising Hamiltonian.
  The ground state of the system is probably the
most important state and through the study of those states we acquire
a lot of information about the corresponding system itself. Beside
being the states with the lowest energy we know that the ground states
are associated with zero temperature and that they are related to
fundamental properties of Hamiltonians. However, there are other
states which are equally relevant  for the physical description of the system.
The temperature of a system we measure in our laboratories is always
non-zero. Keeping in mind the third law
of thermodynamics and the impossibility of reaching the absolute zero
temperature we can conclude that in practice
 there is always a non-zero
probability for finding the system under study in one of the excited
states. Of course, the probability depends on the temperature but as
far as the temperature is non-zero, no matter how big the gap in the
energy between the ground state and the first excited state is, the
probability is non-zero as well. Consequently, it is interesting to
study entanglement in systems in thermal equilibrium, i.e. in their
``natural'' state, and investigate the dependence of the entanglement on
temperature.

The density operator corresponding to a thermal state of a quantum
system at the temperature $T$ is usually given by the relation
\begin{equation}
\label{thermalstate}
\rho (T) = \sum_{i} w_i |e_i \rangle \langle e_i| \; ,
\end{equation}
where $|e_i \rangle $ is an energy eigenstate (eigenstate
of the Hamiltonian $H_3$), $w_i$ are weights or probabilities defined as
\begin{equation}
\label{pwi}
w_i= K e^{ -E_i/T} \; ,
\end{equation}
where we assume the Boltzmann constant to be equal to unity
and we sum over all energy eigenstates. The
constant $K$ in Eq.~(\ref{pwi}) is a normalisation  so the sum of  probabilities $w_i$ equals to unity
\begin{equation}
\sum_i w_i =1 \; .
\end{equation}

\begin{figure}
\includegraphics[width=7cm]{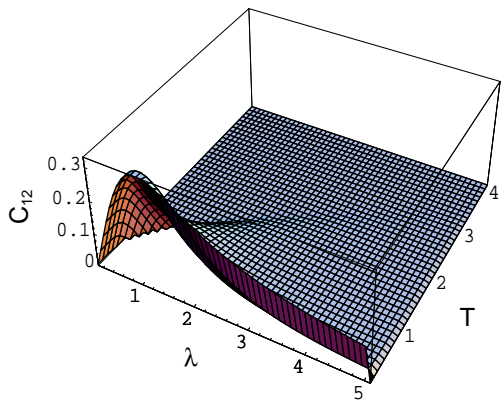}
\caption{The bipartite entanglement between the 1st and the 2nd qubit as
a function of the
temperature $T$ and the parameter $\lambda$. The system of three qubits is in the
thermal state Eq.~(\ref{thermalstate}).}
\label{obr-n3gs}
\end{figure}
In Fig.~\ref{obr-n3gs} we have plotted the entanglement between the
first and second qubit in a three-qubit system at temperature $T$.
 Repeating the same line of arguments, taking
into account the symmetry of the Gibbs state at the temperature $T$,
we know that $C_{12} = C_{13} = C_{23}$ and Fig.~\ref{obr-n3gs} shows us
the dependence of the entanglement on temperature for an arbitrary pair
of qubits. For nearly
  zero values of the temperature the entanglement behaves in a
similar way as in the case of the system in the ground state.
Increasing the temperature the two qubits become
less and less entangled. In the high temperature limit the
entanglement is practically equal to zero. It has a very simple
explanation. If the temperature is high enough all probabilities $w_i$
are almost equal and the state of the system $\rho$ is proportional to
the identity operator~\footnote{The state of the system is not
  proportional to the identity but it is a state which is very close
  to the total mixture.}, i.e. it is the total mixture. Consequently, the state
of an arbitrary pair of qubits is proportional to the identity as well
and the two qubits are not entangled.

In our case an increase of the temperature always causes degradation
of the entanglement. Thus we may conclude that to maximalize the
entanglement it is convenient to keep the temperature as low as
possible. It follows that under certain conditions, one way of
increasing the entanglement can be lowering of the temperature.
At the end let us note that there are quantum models
where an increase of the temperature can cause an increase of
entanglement (see for instance Ref.~\cite{ArnesenBV01}).


\subsection{Quantum entanglement at  $\lambda=1$}
\label{s6.2.4}

Performing an analysis of the entanglement
  for the whole set of eigenstates of the Hamiltonian $H_3$, we have found one particular
    eigenstate with rather peculiar
  behaviour of the entanglement. Namely,
  the entanglement of this state as a function of the parameter
  $\lambda$ is non-analytic at the point $\lambda=1$.

  Let us remind the reader that at the end of Section~\ref{s6.2.1} we
  have mentioned a level crossing. That is, at the point $\lambda=1$
  there is an energy level crossing and one of the non-degenerate
  eigenstates becomes degenerate. What is remarkable is the fact that
  the eigenstate with non-analytic behaviour of entanglement is the
  same state mentioned in Section~\ref{s6.2.1} in connection with the
  level crossing. However, while we have
    discovered the state through our analysis of
    the entanglement of the eigenvectors for three qubits, in
  the general case of arbitrary odd number  of qubits we have followed the reverse path : we have identified
  the state by exploiting the level
  crossing at the point $\lambda=1$~\footnote{Let us note that for $N
    > 3$ qubits there is more than one level crossing.}.

\begin{figure}
\includegraphics[width=7cm]{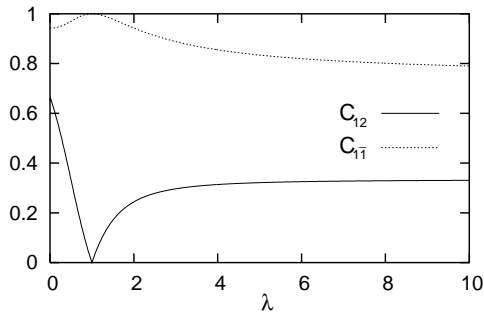}
\caption{The entanglement in the X-state of the Ising chain with three
qubits $N=3$. In the figure we show  two different types of
entanglement as a function of
 the parameter $\lambda$: the bipartite entanglement between the
first and the second qubit $C_{12}$ (solid line) and
the bipartite entanglement
between the first and remaining two qubits $C_{1\bar{1}}$ (dotted line).}
\label{obr-n3xs}
\end{figure}

In what follows we will denote
  the state exhibiting this very intriguing  behaviour as the
  ``{\em X-state}'' (since it exhibits e{\it X}treme entanglement around $\lambda=1$ - for details see
  Sec.~IV.B). In our earlier
    notation, it is the eighth state $|e_8 \rangle$. To remind the
  reader the state has the following form:
\begin{equation}
\label{pstate}
|X \rangle \equiv | v_8 \rangle = K_8 \left
[\frac{1-E_8 - 2 \lambda}{\lambda} |000 \rangle + |011 \rangle + |101 \rangle + |110 \rangle \right ] \; ,
\end{equation}
where $K_8$ is a normalisation constant and $E_8$ is the energy
corresponding to the eigenstate $|e_8 \rangle = | X \rangle$. The
X-state is a non-degenerate eigenstate of the Hamiltonian
(except for a finite number of values of $\lambda$) and thus shares
all symmetries of the Hamiltonian $H_3$ in the same way as the ground
state. That is $C_{12}=C_{13}=C_{23}$ and
$C_{1\bar{1}}=C_{2\bar{2}}=C_{3\bar{3}}$. The bipartite entanglement
between the first and second qubit $C_{12}$ and between the first
qubit and remaining two qubits $C_{1\bar{1}}$ are shown in
Fig.~\ref{obr-n3xs}.
We from from the figure that the concurrence between two qubits in the system
exhibit non-analytical behavior at $\lambda=1$.

Certainly, the reason behind this non-analyticity cannot
 be a phase transition. We know that the Ising
model has a quantum phase transition at the point
$\lambda=1$ but for that the chain must be infinite and the
temperature must be zero \cite{Sagdev}. That is, we can
observe a phase transition only if there is an infinite number of
particles in the chain and the system must be in the ground state.
From this point of view there cannot be a direct link between the
found non-analyticity and the phase transition.

The other question is the relation between
the entanglement and a change of symmetries in the system. The change of symmetries of a
Hamiltonian can have a significant effect on the correlation properties
 of the eigenstates. In our case we know that
 the phase transition is accompanied
with a symmetry change at the corresponding point. Similar change of
symmetry is observed even in the case of a finite dimensional Ising chain
at the point of  the level crossing.
This suggests that the change of the
symmetry at the point $\lambda=1$ may in general be reflected
in the behavior of the
entanglement - the quantum part of the correlations.

The  X-state is interesting not only on
account of the non-analyticity but mainly for the fact that it exhibits
remarkable quantum correlations. At
the point $\lambda=1$ the entanglement in an arbitrary pair of qubits
is zero but the entanglement between a given qubit and the remaining
two qubits is maximal (c.f. Fig.~\ref{obr-n3xs}).
If we calculate the entanglement length introduced in
Ref.~\cite{ArnesenBV01}, it is
zero. But we know that each qubit is maximally entangled with the rest
of the system so there is a sort of a ``long range''
entanglement~\footnote{It is rather misleading to talk about a ``long
  range'' entanglement if there are only three qubits in a chain.
  However, the study will be extended to  many-qubit systems with the
  same result which will justify our terminology.}. In other words, as
each qubit is maximally entangled with the rest of the system and the
pairwise entanglement is zero, we have an intrinsic
multi-partite entanglement. Moreover, since
the system consists of only three qubits the only possible
multi-partite entanglement is a
three-partite entanglement $C_{123}$ and at
the point $\lambda=1$ the three-partite
entanglement reaches its maximal possible value.

Tu sum up we can conclude that in the case of the finite dimensional Ising model,
 there is an energy eigenstate,
  the X-state, for which entanglement exhibits a rather special
  behaviour at the point, where the infinite Ising chain has a
  phase transition.

\section{General case of $N$ spins}
\label{s6.3}

So far we have considered the particular case of
three qubits. Despite its simplicity  the case of three qubits shares many
features of the general case of a chain with
an arbitrary number of qubits. This has helped us to formulate basic theorems
and to identify states which are
particularly interesting  with respect to the
entanglement.

Let us consider a chain with $N$ qubits
where $N$ is arbitrarily large. The Hilbert space of $N$ qubits is
$2^N$-dimensional and the corresponding Hamiltonian $H_N$ has $2^N$
eigenvectors and eigenvalues. Despite the very possibility to
calculate any eigenvector or eigenvalue (recall that the model under
study is exactly solvable) it is not feasible to perform the
calculation for all eigenvectors (eigenvalues) and
to analyse them afterwards. Therefore, we have
used the results of the previous section and beside the ground state
as an important state, we have analysed the X-state. Of course, prior to that we
have to find or identify the X-state among $2^N$ eigenstates of
the Hamiltonian $H_N$. At this point we can take the advantage of our
  detailed knowledge on the spectrum we have at hand, as the level
crossing at the point $\lambda=1$ studied in the previous section is
crucial in identifying the sought state.

Our main goal is to analyse the entanglement properties of the states under consideration.
In addition to the dependence on the
number of qubits $N$ in the chain we will also consider the dependence
on the ``distance'' of qubits. For $N >3$ there are more possibilities how to create pairs of
qubits and beside the nearest neighbours, a pair can be created from
the next nearest neighbours and etc. Since the Ising model
is not permutationally invariant unlike in the special case
studied so far, we can expect that the entanglement will vary
with the distance between qubits.

\subsection{The ground state}
\label{s6.3.1}

The ground state of the system for different values of $N$
can be calculated using the
formalism developed in Appendix~\ref{gensol} and
Appendix~\ref{groundinfty}. Due to the
  complicated form of the state itself, we do not quote the explicit
  expression, it may be found in the above mentioned
  Appendices. In the following, we describe the entanglement
  properties of the state instead.

\begin{figure}
\includegraphics[width=7cm]{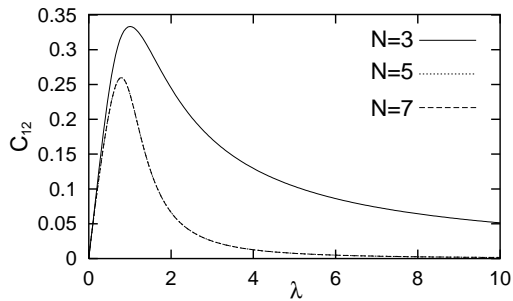}
\caption{The  entanglement between the nearest neighbours as a function of
the parameter $\lambda$ and the number of qubits in the chain $N$. The
system is in the ground state and the number of qubits in the chain is
$N=3,5,7$.}
\label{obr-ngzsqq}
\end{figure}
The entanglement shared between pairs of nearest neighbour qubits in terms of
the concurrence is plotted in Fig.~\ref{obr-ngzsqq} for different
values of $N$. The shape of different curves, corresponding to
different number of qubits in the chain,
is very similar. At the point $\lambda=0$ the values of all curves are zero,
 increasing the parameter $\lambda$ the entanglement (measured in the concurrence) increases and
around the point $\lambda \approx 1$ reaches a maximum.
This maximum depends on the number of
qubits $N$, but with increasing $N$ the concurrence converges to a specific value
that even for $N>5$  is almost constant. Increasing the parameter $\lambda$
further, the entanglement decreases and finally, in the limit
$\lambda \rightarrow \infty $ the entanglement tends to zero.
For $\lambda \to
  \infty$, the ground state is degenerate. Similarly, as in the case
$N=3$, we may choose a particular ground
  state the $\lambda \to \infty$ limit of which becomes the ground
  state of the Hamiltonian for $\lambda \to \infty$ (c.f
  Appendix~\ref{groundinfty})
\begin{equation}
\label{linfN}
|\psi_N \rangle_{\lambda = \infty} = K_N \: \sum_{\{i,j, \ldots\}e }\: |\{i,j, \ldots \} \ldots \rangle \; ,
\end{equation}
where $K_N$ is a normalisation constant, $\{i,j, \ldots \}$ denote
positions of the qubits that are up and $\{i,j, \ldots \}e$ means
summation over all states of the standard basis with an even number of
qubits up. (We use the term ``up'' for a
  qubit if it is in the state $|1 \rangle$ and down if it is in the
state $|0 \rangle$). From the construction of the
state $|\psi_N \rangle_{\lambda = \infty}$, it follows that the entanglement
between arbitrary two qubits is zero while the entanglement between a
given qubit and all remaining qubits is maximal. \\ \mbox{}

 {\em Proof:} The state Eq.~(\ref{linfN}) can
be rewritten into a simpler form using the following recurrence relation:
\begin{eqnarray}
\label{precur}
|\psi_N \rangle_{\lambda = \infty} & = & \frac{1}{\sqrt{2}}
\left  [|0 \rangle |\psi_{N-1} \rangle_{\lambda =\infty}
+ |1 \rangle |\tilde{\psi}_{N-1} \rangle_{\lambda = \infty} \right ] \; ,
\end{eqnarray}
where $| \tilde{\psi}_{N-1} \rangle_{\lambda= \infty} $ has the same
form as $| \psi_{N-1} \rangle_{\lambda = \infty} $ , but instead of
summing over all states with an even number of qubits up we sum over
all states with an odd number of qubits up. With the help of the
Eq.~(\ref{precur}) it is easy to prove the above
statements concerning entanglement. Let $i$ and $j$ denote two arbitrary
but  mutually  different ($i \neq j$) positions of {\em a priori} chosen qubits
 in the chain. Using the relation Eq.~(\ref{precur}) we rewrite the state vector
Eq.~(\ref{linfN}) as follows:
\begin{eqnarray}
|\psi_N \rangle_{\lambda=\infty} = \frac{1}{2} \:
\left  [\: ( \, |00 \rangle_{ij}+|11 \rangle_{ij} \, )
| \psi_{N-2} \rangle_{\lambda = \infty} \right .
\nonumber \\ + \left . \: ( \, |01 \rangle_{ij}
+ |10 \rangle_{ij} \, ) | \tilde{\psi}_{N-2} \rangle_{\lambda = \infty} \right ] \; .
\end{eqnarray}
The reduced density operator $\rho_{ij}$, of the two
qubits at the positions $i$ and $j$ is calculated as
a trace over all  remaining qubits of the density operator
$ \rho = | \psi_N \rangle_{\lambda= \infty} \langle \psi_N |$
of the whole system. The state $\rho_{ij}$ expressed as a
matrix in the basis $\{ | 0 0 \rangle , |01 \rangle,
|10 \rangle, |11 \rangle \}$ reads
\begin{equation}
\label{rhoij}
\rho_{ij} = \left ( \begin{array}{cccc} \frac{1}{4} & 0 & 0 & \frac{1}{4} \\
0 & \frac{1}{4} & \frac{1}{4} & 0 \\
0 & \frac{1}{4} & \frac{1}{4} & 0 \\
\frac{1}{4} & 0 & 0 & \frac{1}{4} \end{array}
\right ) \; .
\end{equation}
The spin-flipped density matrix in this case reads
$\tilde{\rho}_{ij} = \rho$ so that  $\tilde{\rho}_{ij} \rho_{ij} =
\frac{1}{2} \rho_{ij}$. The matrix $\tilde{\rho}_{ij} \rho_{ij}$ has
two eigenvalues that are equal and according to Eq.~\ref{3.4.2.1} the
entanglement shared by the two qubits in the $i$-th and $j$-th
positions is zero.

The density operator of a single qubit at the $i$-th
position $\rho_i$ can be calculated similarly,
\begin{equation}
\label{rhoi}
\rho_i = \left ( \begin{array}{cc} \frac{1}{2} & 0 \\ 0 & \frac{1}{2} \end{array} \right ) \; .
\end{equation}
and  it corresponds to a maximally
  mixed state. Since the whole system is in a pure state,
  the qubit in the $i$-th position is maximally entangled with the
rest of the system. \mbox{} \\

Let us note, that the state given by Eq.~(\ref{linfN}) is a GHZ
  state in the basis built by direct products of eigenvectors
  of $\sigma_x$ of each qubit. Thus the state can be transformed via
  {\it local unitary transformations} into the standard form of the GHZ state
\begin{eqnarray}
\label{pghz}
|GHZ \rangle = \frac{1}{\sqrt{2}} ( |00 \ldots 0 \rangle + |11 \ldots \rangle ) \; .
\end{eqnarray}
This observation provides us with  a very clear explanation of the above mentioned
entanglement properties.  Moreover, it thoroughly confirms the
proposal conjectured at the end of Subsection~\ref{s6.2.2}. For
$\lambda$ being large but not infinite or, equivalently, for a large
coupling constant $C_I$ compared to the absolute value of the magnetic
field $B$, the ground state of the system  exhibits properties close  to the GHZ
state. Taking the parameter $\lambda$ larger and larger the
state is closer and closer to the GHZ state and for a sufficiently
large $\lambda$ we can consider the ground state of the system to be
the GHZ state even for $\lambda$ being large but finite. When the
state of the system is the GHZ state, the entanglement between any
pair of qubits is zero because the reduced density operator describes a separable state
[c.f. Eq.~(\ref{rhoij})]. Further, a reduced density
operator of a subsystem consisting of $n<N$ qubits is also
separable, as one would expect for a GHZ state.  It follows that if we
consider an arbitrary subsystem there is no entanglement: choosing any set of
  $n<N$ qubits, the state of the chosen set is separable. Consequently, the state
  under consideration exhibits only intrinsic $N$-partite entanglement.
Recalling the conjecture from Subsection~\ref{s6.2.2} we can now
confirm the result to be valid for a general case of the $N$-partite chain as well.

Finally, let us note that the ground state (or equivalently the GHZ
state) has a long range entanglement. The
$N$-partite entanglement is certainly
of a long range  since it concerns all
qubits in the chain, though, for instance, the entanglement length defined in
Ref.~\cite{ArnesenBV01} is zero.

\begin{figure}
\includegraphics[width=7cm]{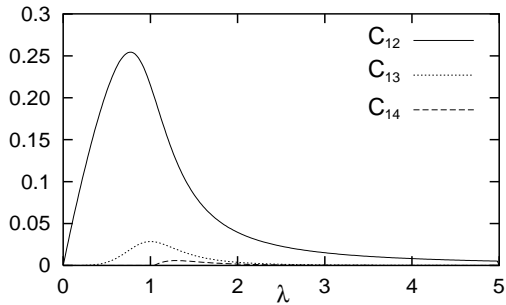}
\caption{The
bipartite entanglement shared between qubits, expressed in terms of the
concurrence $C_{ij}$ where $i$ and $j$ label positions of two qubits, as
a function of
the parameter $\lambda$ and the distance between the qubits.
The system (chain) of seven qubits is in the ground state.}
\label{obr-n7zsqq}
\end{figure}

We have also studied how the entanglement depends
on the distance between an arbitrary pair of qubits qubits, see Fig.~\ref{obr-n7zsqq}.
The
farther the two qubits are
they are less entangled. It means that
with increasing the distance (specified by positions of qubits, i.e. the distance
between qubits $j_1$ and $j_2$
is represented by the difference $|j_1-j_2|$)
between qubits the entanglement converges to zero. Besides, the peak, or the
point where the entanglement is maximal, is shifted to larger
$\lambda$'s (see Fig.~\ref{obr-n7zsqq}).

\subsection{General  X-state }
\label{s6.3.2}

In the case of the chain composed of
 three qubits  we have found a
  particular eigenstate of the Hamiltonian $H_3$, the
  X-state, with a non-analytic behaviour of entanglement.
The question naturally arises, whether there exists such a state
  in the case of more than three qubits. As the Hamiltonian $H_N$ has $2^N$ eigenvectors
  and eigenvalues it is impossible to analyse the whole spectrum.
   However, we know that in the three qubit case the
  X-state is interesting not only on the account of entanglement
  but also because of  energy-level crossing.
  It is a non-degenerate eigenvector of
  $H_3$ apart from a single point $\lambda=1$ where there is a level
  crossing. Our knowledge of the level crossing can be successfully exploited in identifying the
  X-state for an arbitrary $N$. We have found that
  for $N$ odd there is a level crossing at the point
  $\lambda=1$~\footnote{For $N$ large there are several level
    crossings for different values of the parameter $\lambda$. The
    one we are interested in is
    at $\lambda=1$.} and one of non-degenerate
  eigenvectors becomes degenerate. Let us note that in the case $N=3$
  the X-state corresponds to the non-degenerate eigenvector while
  the other energy level is degenerate and corresponds to two vectors.
  The situation we have now is similar. There is a level crossing of
  two energy levels at the point $\lambda=1$. One of them is a
  non-degenerate energy level; in what follow we will call the state
  corresponding to that level as the X-state, and the other energy
  level is degenerate and there are $2^n$ (where $N=2n +1$)
  eigenvectors corresponding to that level (for proofs and more
  details see Appendix~\ref{pstatea}).

  Having successfully identified the
  X-state, we can analyse its entanglement properties. In contrast with the case of $N=3$ qubits, for $N>3$
  bipartite entanglement (the concurrence) as a function of the
  parameter $\lambda$ is analytic.  In order to see this we need to know the form of the
  X-state. It is derived in
    Appendix~\ref{gensol}, we quote only the result for $\lambda=1$ here
    (see Appendix~\ref{pstatea} and also Ref.\cite{Fort})
\begin{equation}
\label{def-p}
|X \rangle_{\lambda=1} = \sum_{\{i,j, \ldots\}_e } | \{ i,j,k,\ldots \} \rangle (-1)^{ \sum_{i>j} d(i,j)} \; ,
\end{equation}
where $\{i,j, \ldots \}_e$ denote a sum over all sets of indices with
an even number of indices in each set, the letters $i,j,\ldots$ in a
single set denote positions of qubits in the chain that are
up and $|\{i,j, \ldots
\} \rangle$ is the corresponding state vector, $d(i,j)$ is a distance
between qubits on the $i$-th and $j$-th positions defined below,
while the sum over $i<j$ means that we sum over all
pairs of qubits counting only once the cases with switched positions
of the qubits. The distance $d(i,j)$ of the two qubits is defined as
the shortest path on the ring that brings us from the qubit on the
$i$-th positions to the qubit on the $j$-th position.  In Appendix~\ref{pstatea} we
present a complete proof that the state Eq.~(\ref{def-p}) is an
eigenstate of the Hamiltonian $H_N$ for $\lambda=1$.
The Appendix~\ref{pstatea} also contains several additional
  proofs and more details on the $X$-state.

  The reduced density operator of two qubits on the $i$-th and $j$-th
  positions has been calculated from the state Eq.~(\ref{def-p}) by
  tracing over degrees of freedom of the remaining qubits. The density
  operator expressed in the basis $\{ |00\rangle, |01 \rangle, |10
  \rangle, |11 \rangle\}$ reads
\begin{eqnarray}
\label{rhoijp}
\rho_{ij} = \left ( \begin{array}{cccc} \frac{1}{4} & 0 & 0 & 0 \\
0 & \frac{1}{4} & 0 & 0 \\
0 & 0 & \frac{1}{4} & 0 \\
0 & 0 & 0 & \frac{1}{4} \end{array} \right ) \; .
\end{eqnarray}
In the same way the density operator of a single qubit in the $i$-th
position~\footnote{The operator can be calculated as a trace of the
  density matrix Eq.~(\ref{rhoijp}) over the degrees of freedom of the
  qubit on the $j$-th position or directly from the state
  Eq.~(\ref{def-p}).} expressed in the one-qubit basis $\{|0 \rangle,
|1 \rangle\}$ is
\begin{eqnarray}
\label{rhoip}
\rho_i = \left ( \begin{array}{cc} \frac{1}{2} & 0 \\
0 & \frac{1}{2} \end{array} \right ) .
\end{eqnarray}

With the help of the density matrix Eq.~(\ref{rhoijp}) we have
calculated bipartite entanglement between qubits on the $i$-th and
$j$-th positions while the density matrix Eq.~(\ref{rhoip}) has been
used for the calculation of entanglement between a qubit on the $i$-th
position and the rest of the system (all remaining qubits). Since the
density matrix Eq.~(\ref{rhoijp}) is a complete mixture,  there is no bipartite
entanglement between any two qubits. On the other hand as
  Eq.~(\ref{rhoip})  describes a complete mixture and the whole
  system is in a pure state, entanglement between a given qubit and
remaining qubits is obviously maximal: the tangle is
  equal to one. As the eigenvalues of the
density matrices $\rho_{ij}$ and $\rho_i$ are continuous functions of the parameter
$\lambda$ it is easy to check that both the entanglement shared
between qubits and the entanglement between a given qubit and the rest
of the system are continuous functions of the parameter $\lambda$
around the point $\lambda=1$. Consequently, the point $\lambda=1$ is
not a point of non-analytical behaviour of entanglement anymore.
Moreover, the state Eq.~(\ref{def-p}) is not equivalent to a GHZ
state. In other words there does not exist a local unitary
transformation which would transform the state Eq.~(\ref{def-p}) into
the GHZ state in Eq.~(\ref{pghz}).  However, we have
found that the X-state Eq.~(\ref{def-p}) has the following
remarkable property:
\mbox{} \\

{\bf Theorem 1 }{\em Let $N=2n+1$ denote the total number of qubits forming the Ising
chain where $n$ is an integer and let the system of $N$ qubits be in the  X-state Eq.~(\ref{def-p}).

The density matrix of any sequence of $n$ neighboring qubits is
\begin{equation}
\label{pnnq}
\rho = \frac{1}{2^n} \openone_n \; ,
\end{equation}
where $\openone_n $ is the identity operator acting in the
$2^n$-dimensional Hilbert space of $n$ qubits. } \\ \mbox{} \\
An important consequence of the theorem is the fact that if the system
is in the X-state then any set of neighboring qubits is perfectly
(maximally) entangled with the rest of the system. Consider an
arbitrary set of $n$ neighboring qubits. The reduced density operator
of the $n$ qubits is Eq.~(\ref{pnnq}). Using the entropy Eq.~(\ref{entropy})
as a measure
of bipartite entanglement for pure states
\begin{eqnarray}
\label{ent}
S = - \sum_{2^n} \frac{1}{2^n} \log \frac{1}{2^n} = n \; \log 2 \; ,
\end{eqnarray}
we can see that the set of $n$ qubits is maximally entangled with the
remaining qubits. That is, if a system of $N=2n+1$ qubits is in the
X-state and we choose $n$ neighboring qubits then according to
Eq.~(\ref{ent}) we know that the $n$ qubits are perfectly entangled
with the remaining $n+1$ qubits. Moreover, if we choose a subset of
say $k$ qubits from the set of $n$ neighboring qubits then the state
of the $k$ qubits is
\begin{equation}
\rho = \frac{1}{2^k} \: \openone_k \; ,
\end{equation}
so the $k$ qubits are perfectly entangled with the rest of the system
(all remaining $N-k$ qubits). To sum up the X-state is a highly
entangled state and consequently it is a good candidate for a quantum
communication between many parties. A rather simple protocol that can serve as an example of
  its applications described in the next section.

\subsection{Controlling  distribution of entanglement in the X-state}

We will present a simple example how the X-state can be exploited
for a communication or a secret key distribution in a situation
when bipartite entanglement between many parties is needed. The X-state
with its remarkable properties can be considered to be a very
good resource of communication as any set of $n$ neighboring qubits is
maximally entangled with the rest of the system.

First, imagine that  $n$ neighboring qubits belong to Alice
and the rest ($n+1$ qubits) belongs to Bob. Moreover, let us assume
that Alice and Bob want to exploit the entanglement of the X-state for
their protocol. But unfortunately, their protocol is designed for
qubits, that is to say it uses pairs of maximally entangled qubits. We
have shown that the density operator of any pair of qubits is
proportional to the identity and thus the two qubits cannot be
entangled. It means that Alice and Bob cannot take any two qubits and
use them for their protocol. But we know that the $n$ neighboring
qubits which belongs to Alice are maximally entangled with Bob's
qubits as the entropy in Eq.~(\ref{ent}) equals to $n \: \log 2$.
Such amount of entanglement corresponds to $n$ pairs of maximally
entangled qubits. Therefore one may
  ask whether they are able to create $n$ pairs of maximally
entangled qubits only by performing {\em local} (though multi-qubit) unitary transformations
${\bf U}_A^{(n)}$ and ${\bf U}_B^{(n+1)}$ on their respective qubits.
The answer is positive.

Consider a state of $2n+1$ qubits with $n$ pairs of maximally
entangled qubits and let  Bob's last qubit be in the state $|0
\rangle$ (We know that unlike Alice, Bob has got $n+1$ qubits.):
\begin{eqnarray*}
| \Xi \rangle = \left ( \bigotimes_n \frac{1}{\sqrt{2}}
( |00 \rangle_{AB} + |11 \rangle_{AB} ) \right ) \otimes |0 \rangle_B \; .
\end{eqnarray*}
Now let us reorder the qubits in such a way that the first $n$ qubits belong
to Alice and the remaining $n+1$ qubits belong to Bob. We need to do
that because Alice possesses $n$ {\em neighboring} qubits.
\begin{eqnarray}
\label{4.10}
| \Xi \rangle = \sum_{\{ i,j, \ldots\}}
| \{i,j, \ldots \} \rangle_A \otimes \frac{1}{2^{n/2}} | \{ i,j, \ldots \} \rangle \otimes | 0 \rangle
\; ,
\end{eqnarray}
where a set of indices $ \{i,j, \ldots \}$ denotes positions of qubits
up in the standard basis vector and we sum over all possible sets of
indices. Now we want to find a local unitary transformation ${\bf U} =
{\bf U}_A^{(n)} \otimes {\bf U}_B^{(n+1)}$ such that the state $ | \Xi
\rangle$ transforms into X-state. It follows from Eq. (\ref{C13})
that it is enough to consider the unitary transformation of the form
${\bf U} = \openone_A \otimes {\bf U}_B^{(n+1)}$ where
\begin{eqnarray*}
{\bf U}_B^{(n+1)}: | \alpha_{\{ i,j \ldots \} } \rangle_{\bar{O}}
\rightarrow \frac{1}{2^{n/2}} | \{ i,j, \ldots \} \rangle \otimes | 0 \rangle \; ,
\end{eqnarray*}
and the states $| \alpha_{\{ i,j \ldots \} } \rangle_{\bar{O}}$ are
defined in the Appendix~\ref{pstatea}.
After Bob has performed the unitary operation Alice and Bob share $n$
pairs of maximally entangled qubits and they can begin
  with their original protocol. This simple example
illuminates the remarkable properties of the X-state and its
convenient form since only Bob has to perform the local unitary
operation.

The situation becomes even more interesting if we replace Alice with
$n$ parties $\{A_1, \ldots A_n \}$. Now, Bob communicates with $n$
different parties. By performing a local operation he can decide which
of his qubits is entangled with a given partner $A_j$. Let us stress
that this is only a simple example and more sophisticated protocols
are the topic of current research.

\section{Summary}
\label{s6.4}

We have performed detailed analytical calculations
  concerning stationary states of a finite-size Ising chain with
  cyclic boundary conditions, and their entanglement properties. We
  have put a special emphasis to a kind of description of multi-partite
  entanglement.

The primary motivation of our investigation has been an attempt to
illuminate the Ising model using tools of quantum information theory. In
addition we were studying a possibility whether some properties of the Ising
model can be used as a resource of quantum information processing /
communication. For this purpose, one of the criteria that should be met is
that of the possibility of preparing the system in a suitable initial state.
As physical systems tend to occupy their ground states, it is advantageous
if the ground state is a suitable initial state for some purposes We have
shown that by adjusting the external magnetic field the ground state of the
model considered is the well known GHZ state used in several quantum
information processing schemes.  Consequently, with the ground state of the
system well known and having particularly nice properties makes the Ising
model a good candidate as a resource for a quantum information processing.

This result also demonstrates the usefulness of the approach to finding an
entangled state with pre-defined multi-partite entanglement properties by
finding the ground state of a suitably chosen Hamiltonian. The ground state
of the Ising model for certain values of the parameter $\lambda$ is a very
specific state - the GHZ state. The GHZ state has the property that the
entanglement between any set of $n$ qubits where $n<N$ is zero while the
$N$-partite entanglement peaks reaching the maximum possible value in the
limit $\lambda \rightarrow \infty$. It means that in the limit $\lambda
\rightarrow \infty$ the ground state of the Ising model maximises the
$N$-partite entanglement and the state of $N$
qubits with maximum $N$-partite entanglement can be found as a ground state
of the Ising model with $N$ qubits in the chain.

Regarding entanglement properties, not only the ground state of the Ising
Hamiltonian is found to be interesting. For instance, we have discovered a
very specific eigenstate of the Hamiltonian, the X-state. The X-state is
{\em strongly} (extremely) entangled as every set of $k \leq n$ neighboring
qubits (where the total number of qubits is $N=2n+1$) is maximally entangled with the remaining qubits. An
important message of our results is, that multi-partite entanglement plays a
crucial role in the understanding of exactly solvable models of quantum
statistics. We have also presented a simple example  to demonstrate
the usefulness of such state in quantum communication.

The X-state is  identified via a certain crossing of energy levels at
 $\lambda=1$, where a phase transition occurs in the
thermodynamic (infinite-qubit) limit. Consequently, there might be some
connections between a functional dependence of the entanglement as a
function of $\lambda$ and the point of a phase transition. One important
lesson one can learn from our investigation is that higher energy
eigenstates of the Ising Hamiltonian might carry non-trivial information
about quantum correlations of the system under consideration.


This was work supported in part
by  the European Union projects QGATES, QUPRODIS,   and
CONQUEST and by  the Slovak Academy of Sciences.
We would like to thank Matyas Koniorczyk for his helpful comments.
VB thanks Michael Nielsen for stimulating discussions and hospitality
during his visit to the Queensland
University in 2000 when this work has been done.

\appendix

\section{Exact solution of the $1D$ Ising chain}
\label{gensol}

The Ising model as  one of the simplest exactly solvable models
has been studied extensively in the literature and  there is  a
chapter on the Ising model in almost every textbook on solid state
physics with  exactly solvable models. The derivation here will mainly,
 up to a few minor deviations, follow Ref.~\cite{Izumov}.
 The reason why we summarise  the
derivation here is the fact that we present a complete solution
together with the exact form of the eigenvectors  usually
omitted in the literature and also to keep our discussion self-content.

Consider   the  Hamiltonian Eq.~(\ref{6.1.3}) with the cyclic
conditions Eq.~(\ref{6.1.2}). Our task is to find the eigenvalues and
eigenvectors of this Hamiltonian. As a first step we perform
several transformations. This is a bit technical part however it
needs to be included in the derivation to keep it transparent.

\subsection{Transition from variables $ \sigma^{\alpha} $ to $ \sigma^{\pm} $}
Firstly we  introduce new variables $\sigma^{\pm}_n$ where
\begin{eqnarray*}
\sigma^{\pm}_n & = & \frac{1}{2} ( \sigma^{x}_n \pm i \sigma^{y}_n) \; . \\
\end{eqnarray*}
The Hamiltonian Eq.~(\ref{6.1.3}) after the first transformation reads
\begin{eqnarray*}
{H} & = & -
\lambda \sum_{n=1}^{N} (\sigma_{n}^{+} + \sigma_{n}^{-}) ( \sigma_{n+1}^{+} + \sigma_{n+1}^{-}) \\
& & +  2 \sum_{n=1}^{N} (\sigma_{n}^{+} \sigma_{n}^{-} - \frac{1}{2} {\bf 1}) \; .
\end{eqnarray*}

\subsection{Jordan-Weyl transformation}
Secondly, we introduce fermionic variables $c_n^{\dagger}$ and
$c_n$ such that
\begin{eqnarray*}
c_{n} & = & \exp(i \pi \sum_{j=1}^{n-1} \sigma_{j}^{+} \sigma_{j}^{-}) \sigma_n^{-} \; ,\\
c_{n}^{\dagger} & = & \exp(i \pi \sum_{j=1}^{n-1} \sigma_{j}^{+}
\sigma_{j}^{-}) \sigma_n^{+} \; .
\end{eqnarray*}
The variables $c_n^\dagger$ and $c_n$  satisfy the
anti-commutation relations. The form of the Hamiltonian after the
second transformation is following
\begin{eqnarray*}
{H} & = & - \lambda \sum_{n=1}^{N-1} \left ( (c_n^{\dagger} - c_n) ( c_{m+1}^{\dagger} + c_{m+1}) \right . \\
& & \left .  -  (-1)^{\hat{N}} (c_N^{\dagger} -c_N) (c_1^{\dagger} + c_1) \right ) \\
& & + 2 \sum_{n=1}^{N} \left ( c_n^{\dagger} c_n - \frac{1}{2} \right )  \; ,
\end{eqnarray*}
where \( \hat{N} = \sum_{n=1}^{N} c_n^{\dagger} c_n \)  can be
interpreted as the operator of number of fermions. The operator
$(-1)^{\hat{N}} $  commutes with the Hamiltonian $ H $. For that
reason it is possible to choose common eigenvectors of the
Hamiltonian and the operator $(-1)^{\hat{N}}$. The eigenvalues of
the operator $\hat{N}$ are even or odd. If an  eigenvector of the
operator $\hat{N}$ corresponds to an {\em even}  eigenvalue then
the operators $c_{N+1}$ and $c_{N+1}^\dagger$ satisfy the condition
\begin{eqnarray*}
c_{N+1}^{\dagger} & = & - c_1^{\dagger} \mbox{\hspace{1in}} anticyclic \; cond.\\
c_{N+1} & = & - c_1 \; .
\end{eqnarray*}
On the other hand if an eigenvector of the operator $\hat{N}$ corresponds to
an {\em odd} eigenvalue, the operators $c_{N+1}$ and $c_{N+1}^\dagger$ are
defined as
\begin{eqnarray*}
c_{N+1}^{\dagger} & = & c_1^{\dagger} \mbox{\hspace{1in}} cyclic \;  cond.\\
xc_{N+1} & = & c_1 \; .
\end{eqnarray*}
The advantage of the previous choice  is that the Hamiltonian has the same form in both cases
\begin{eqnarray*}
{H} & = & - \lambda \sum_{n=1}^{N} \left ( (c_n^{\dagger} - c_n) ( c_{n+1}^{\dagger} + c_{n+1}) \right  ) \\
& & + 2 \sum_{n=1}^{N} \left ( c_n^{\dagger} c_n - \frac{1}{2} \right ) \; .
\end{eqnarray*}

\subsection{Momentum representation}

The  last transformation is a transition to the momentum
representation and specified by the variables $\eta_q^\dagger$ and
$\eta_q$. The fermionic operators $c_n^\dagger$ and $c_n$ and the
new operators $\eta_q^\dagger$ and $\eta_q$ are related as
\begin{eqnarray*}
c_m & = & \frac{1}{\sqrt{N}} \sum_q e^{i q m } \eta_q \; , \\
c_m^{\dagger} & = & \frac{1}{\sqrt{N}} \sum_q e^{-i q m} \eta_q^{\dagger} \; ,
\end{eqnarray*}
where $q$ depends on the  boundary conditions.
For cyclic boundary conditions ($ \hat{N}$  is odd) we have
\begin{eqnarray*}
e^{iqN} & = & 1 \; ,
\end{eqnarray*}
so that
\begin{eqnarray*}
q & = & \frac{\pi}{N} 2 l \; \; \; \; l = 0, 1, 2 \ldots (N-1) \; .
\end{eqnarray*}
For  anti-cyclic boundary conditions ($ \hat{N}$  is even)
\begin{eqnarray*}
e^{iqN} & = & - 1 \; ,
\end{eqnarray*}
we have
\begin{eqnarray*}
q & = & \frac{\pi}{N} (2l+1) \; \; \; \; l = 0, 1, 2 \ldots (N-1) \: .
\end{eqnarray*}
The Hamiltonian $H$ has again the same form in both cases and is given by
\begin{eqnarray*}
{H} & = \sum_q & \left [ - \lambda \cos{q} ( \eta_q^{\dagger} \eta_{2 \pi-q}^{\dagger} + 2 \eta_q^{\dagger} \eta_q - \eta_q \eta_{2\pi -q} ) \right . \\
& & \left .  -\lambda  i \sin{q} ( \eta_q^{\dagger} \eta_{2\pi -q}^{\dagger} + \eta_q \eta_{2\pi-q} ) \right . \\
& & \left . + 2\: ( \eta_q^{\dagger} \eta_q - \frac{1}{2}) \right ] .
\end{eqnarray*}
It is important to note  that the operator $\hat{N}$ can easily be
rewritten using the new variables $\eta_q$ and $\eta_q^{\dagger}$
\begin{eqnarray}
\label{pahq}
\hat{N} = \sum_n c_n^{\dagger} c_n = \sum_q \eta_q^{\dagger} \eta_q  \; .
\end{eqnarray}
It follows that the number of fermions  described with $c_m$ and $c_m^{\dagger}$ is the same as the number of other fermions described with the operators $\eta_q$ and $\eta_q^{\dagger}$.

\subsection{Eigenvalues and eigenvectors of $ {H}_q$ }

In what follows we calculate  eigenvalues and eigenvectors of the
Hamiltonian $H$. Firstly, it is convenient to reorder the
contributions to the sum Eq.~(\ref{pahq}). Specifically,  choose
one $q$, and let us define $ {H}_q$ (we note, that it is not
possible to perform this step when $q=0$ or $q=\pi$. In that case
see the end of this paragraph.)
\begin{eqnarray*}
{H}_q  & = & - \lambda \cos{q} (2 \eta_q^{\dagger} \eta_q + 2 \eta_{2\pi-q}^{\dagger} \eta_{2\pi-q}) \\
& & - \lambda i \sin{q} (2 \eta_q^{\dagger} \eta_{2\pi-q}^{\dagger} +2 \eta_q \eta_{2\pi-q}) \\
& & + 2 (\eta_q^{\dagger} \eta_q + \eta_{2\pi-q}^{\dagger} \eta_{2\pi-q} - 1 ) \; .
\end{eqnarray*}
The Hamiltonian $H$ is a sum of  $ {H}_q$
\begin{eqnarray*}
H = \hat{\sum_q} {H}_q
\end{eqnarray*}
where we sum over half of the q's since $H_q = H_{2 \pi -q}$.
The point in rewriting the Hamiltonian
as a sum of $ {H}_q$ is that it is easy to calculate the eigenstates and eigenvalues of $ {H}_q$.
For each $ {H}_q$ we define the following basis
\begin{eqnarray*}
| \Phi_1 \rangle & = & \eta_q^{\dagger} |0> \; ,\\
| \Phi_2 \rangle & = & \eta_{2\pi-q}^{\dagger} |0> \; ,\\
| \Phi_3 \rangle & = & |0> \; ,\\
| \Phi_4 \rangle & = & \eta_{2\pi-q}^{\dagger} \eta_q^{\dagger} |0>.
\end{eqnarray*}
In this particular basis the Hamiltonian $ {H}_q$ is a $4 \times 4$ matrix
\[ \left ( \begin{array}{cccc}
- 2 \lambda \cos{q} & 0 & 0 & 0 \\
0 & -2 \lambda \cos{q} & 0 & 0 \\
0 & 0 & -2 & -2 \lambda i \sin{q} \\
0 & 0 & 2 \lambda i sin{q} & 2 - 4 \lambda \cos{q} \\
\end{array} \right ) \; ,\]
and the four eigenvalues of $ {H}_q$ are
\begin{eqnarray*}
a_1 & = & -2 \lambda \cos{q} \; ,\\
a_2 & = & -2 \lambda \cos{q} \; ,\\
a_3 & = & 2 ( - \lambda \cos{q} + \sqrt{(\lambda-1)^2 + 2 \lambda (1-\cos{q})}) \; , \\
a_4 & = & 2 ( - \lambda \cos{q} - \sqrt{(\lambda-1)^2 + 2 \lambda (1-\cos{q})})  \; .\\
\end{eqnarray*}
The  eigenvectors corresponding to the eigenvalues $a_1$, $a_2$, $a_3$ and $a_4$
in the defined basis reads
\begin{eqnarray*}
|a_1 \rangle  & = & | \Phi_1 \rangle \; ,\\
|a_2 \rangle  & = & | \Phi_2 \rangle \; ,\\
|a_3 \rangle  & = & d_3 | \Phi_3 \rangle + e_3 | \Phi_4 \rangle \; ,\\
|a_4 \rangle  & = & d_4 | \Phi_3 \rangle + e_4 | \Phi_4 \rangle \; ,
\end{eqnarray*}
where
\begin{eqnarray*}
d_j & = & - \frac{2 \: \lambda \: i \: \sin{q}}{a_j +2} \; e_j \; ,\\
e_j & = & \frac{1}{\sqrt{1+ \frac{4 \lambda^2 (\sin{q})^2}{(a_j + 2)^2}}}.
\end{eqnarray*}
At this moment we postpone the derivation of the  eigenvectors and
eigenstates of the Hamiltonian $H$ to the next paragraph and
instead we discuss the cases when $q=0$ (or
$q=\pi$)~\footnote{Actually, we discuss the case $q = 0$ while in
the brackets we present the corresponding expressions for
$q=\pi$.}. The form of operator  $ {H}_0$ ( $H_{\pi}$ ) is
\begin{eqnarray*}
 {H}_0 & = & - 2 \: \lambda  \: \eta_0^{\dagger} \: \eta_0
+ 2 \: \eta_0^{\dagger} \: eta_0 - 1 \hspace{0.5in} q=0 \\
( \; \;  \hspace{0.15in}  {H}_{\pi}  & = & +2 \: \lambda \: \eta_{\pi}^{\dagger} \eta_{\pi} + 2 \: \eta_{\pi}^{\dagger} \eta_{\pi} - 1 \hspace{0.47in} q=\pi  \; \; ) .
\end{eqnarray*}
In this case we define the basis as follows
\begin{eqnarray*}
| \Phi_{q=0} \rangle & = & \eta_0^{\dagger} |0> \hspace{0.56in} [ \; \;  | \Phi_{\pi} \rangle = \eta_{\pi}^{\dagger} |0> \; \; ] \\
| \Phi_0 \rangle & = & |0> \hspace{0.69in} [ \; \; | \Phi_0 \rangle = | 0 \rangle \; \; ] \; . \\
\end{eqnarray*}
The operator $H_0$ ($H_{\pi}$ )  in this particular basis is a $2
\times 2$ matrix
\[ \left ( \begin{array}{cc} -1 & 0 \\ 0 & -2 \lambda +1 \end{array} \right )
\hspace{0.7in} \left [ \; \left ( \begin{array}{cc} -1 & 0 \\ 0 & 2 \lambda +1 \end{array} \right ) \; \right ] \; ,\]
The matrix has two eigenvalues
\begin{eqnarray*}
a_1 & = & -1 \hspace{0.98in} [ \; \; a_1 = -1 \; \; ] \; , \\
a_2 & = & -2 \lambda +1 \hspace{0.66in} [ \; \; a_2 = 2 \lambda +1 \; \; ] \; ,
\end{eqnarray*}
with the corresponding eigenvectors given by
\begin{eqnarray*}
|a_1 \rangle  & = & | \Phi_0 \rangle \hspace{0.9in} [\; \; |a_1 \rangle  = | \Phi_0 \rangle \; \; ] \; , \\
|a_2 \rangle  & = & | \Phi_{q=0} \rangle \hspace{0.76in} [ \; \; |a_2 \rangle  = | \Phi_{\pi} \rangle \; \;  ] \; .
\end{eqnarray*}

\subsection{Eigenvalues and eigenvectors of H: Example}

According to the previous paragraph  the Hamiltonian $H$ can be
rewritten as a sum of the operators $H_q$. Since each operator
$H_q$ is a sum of terms with an even number of fermionic operators
and two different $H_q$'s contain different fermionic operators we
conclude that the operators $H_q$ commute. Essentially it means
that they have common eigenvectors. As an example let us consider
the case $N=5$ and let $(-1)^{\hat{N}}$. Then one of the energy
eigenstates is
\[ |e_H \rangle =  \eta_{0}^{\dagger}
\eta_{2 \pi /5 }^{\dagger} \eta_{4 \pi /5}^{\dagger}  |0 \rangle  \]
with  the corresponding eigenvalue
\[ E_H = -1 - 2 \lambda \cos (\frac{2 \pi}{5})  - 2 \lambda \cos ( \frac{4 \pi}{5} ). \]
We have to keep in mind here that not all the combinations are the eigenstates of $H$. For example if $(-1)^{\hat{N}}=-1$  then the state
\begin{equation}
\label{pae1}
 | \varphi \rangle =  \eta_{2 \pi /5 }^{\dagger} \eta_{4 \pi /5}^{\dagger}  |0 \rangle  \; ,
\end{equation}
is not an eigenstate of $H$ even though this state is an eigenvector of $H_0$, $H_{2 \pi /5}$ and $H_{4 \pi / 5}$. The state Eq.~(\ref{pae1}) has even number of fermions and consequently for $\hat{N}$ even the Hamiltonian is not
\[ H = H_{0} + H_{2 \pi /5} + H_{4 \pi /5} \; ,\]
but
\[ H = H_{\pi /5} + H_{3 \pi /5} + H_{\pi} \; .\]
Recall that  $q$'s depend on the boundary conditions which in turn
depend on the number of fermions.

\section{Three spins: Eigenvalues and eigenvectors of the Hamiltonian $H_3$.}
\label{p3vhvv}
For completeness we review the complete spectrum of
the Hamiltonian $H_3$. The Hilbert space
corresponding to the system of three qubits is eight dimensional
and the Hamiltonian $H_3$ has eight eigenvalues,
\begin{eqnarray*}
E_{1,2} & = &  \lambda +1  \; ,\\
E_{3,4} & = &  \lambda -1  \; ,\\
E_5 & = & 1 - \lambda - 2 \sqrt{1 + \lambda + \lambda^2} \; ,\\
E_6 & = & 1 - \lambda + 2 \sqrt{1 + \lambda + \lambda^2} \; ,\\
E_7 & = & -1 -\lambda - 2 \sqrt{1 -\lambda + \lambda^2} \; ,\\
E_8 & = & - 1 - \lambda + 2 \sqrt{1 -\lambda + \lambda^2} \; .
\end{eqnarray*}
The eigenvalues $E_1 = E_2$ and $E_3 = E_4$ are degenerate for all values of parameter $\lambda$ while the other four are (apart from a finite number of points) not. The eigenvectors corresponding to the eigenvalues read
\begin{eqnarray*}
|e_1 \rangle & = & \frac{1}{\sqrt{2}} \left  [|110 \rangle - |011 \rangle \right ] \; ,\\
|e_2 \rangle & = & \frac{1}{\sqrt{2}} \left  [|101 \rangle - |011 \rangle \right ] \; ,\\
|e_3 \rangle & = & \frac{1}{\sqrt{2}} \left  [|100 \rangle - |001 \rangle \right ] \; ,\\
|e_4 \rangle & = & \frac{1}{\sqrt{2}} \left  [|010 \rangle - |001 \rangle \right ] \; ,\\
|e_5 \rangle & = & K_5 \left  [\frac{ E_5 + 1 - \lambda}{E_5 - 3 - \lambda} |111 \rangle + |001 \rangle + |010 \rangle + |100 \rangle \right ] \; ,\\
|e_6 \rangle & = & K_6 \left  [\frac{ E_6 + 1 - \lambda}{E_6 - 3 - \lambda} |111 \rangle + |001 \rangle + |010 \rangle + |100 \rangle \right ] \; ,\\
|e_7 \rangle & = & K_7 \left  [\frac{1-E_7 - 2\lambda}{\lambda} |000 \rangle + |011 \rangle + |101 \rangle + |110 \rangle \right ] \; ,\\
|e_8 \rangle & = & K_8 \left  [\frac{1-E_8 - 2\lambda}{\lambda} |000 \rangle + |011 \rangle + |101 \rangle + |110 \rangle\right ] \; .
\end{eqnarray*}

\section{Ground state  for $\lambda = \infty$}
\label{groundinfty}

The case $\lambda=\infty$
corresponds to the physical situation with  zero magnetic field $B$.
For the given value of the parameter $\lambda$ the Hamiltonian $H_N$ has
the following form:
\begin{eqnarray*}
H = - C_I \sum_{i=1}^N \sigma_i^x \otimes \sigma_{i+1}^x \; .
\end{eqnarray*}
The ground state of the Hamiltonian is degenerate, the
energy of the ground state is $E_g = - N C_I$ and two energy states
related to the energy $E_g = -N C_I$ are
\begin{eqnarray}
\label{2grounds}
|\bar{g}_1 \rangle = | \bar{0} \bar{0} \ldots \rangle \; ;\nonumber \\
|\bar{g}_2 \rangle = | \bar{1} \bar{1} \ldots \rangle \; ,
\end{eqnarray}
where $|\bar{0} \rangle $ and $|\bar{1} \rangle$ are  eigenstates
of $\sigma^x$. Of course not only these two states are
eigenstates of the Hamiltonian.
Any  linear combinations of these states is also an
eigenstate. However, we are searching for  eigenstates that are
the limit of the eigenstates of the Hamiltonian (\ref{6.1.3}) when
$\lambda$ tends to infinity. At this point we use the knowledge of
the exact solution (see Appendix~\ref{gensol}), and more specifically the fact that
any  eigenstate of the Hamiltonian is a linear
combination of standard basis vectors with either even or odd
number of qubits up. (Standard basis corresponds to state vectors that are
eigenstates of all $\sigma_i^z$. The fact that a state is a linear
combination of standard basis vectors with either odd or even number
of qubits up is equivalent to the condition that the state is an
eigenstate of the operator $(-1)^{\hat{N}}$.) Therefore we select two
particular linear combinations, such that they are sums of state
vectors of standard basis with either even or odd number of qubits
up.
The linear combinations of the two vectors in Eq.~(\ref{2grounds}) which  satisfy the condition are
\begin{eqnarray*}
|g_1 \rangle = \frac{1}{\sqrt{2}} (| \bar{0} \bar{0} \ldots \rangle +  | \bar{1} \bar{1} \ldots \rangle ) \; ,\\
|g_2 \rangle = \frac{1}{\sqrt{2}} (| \bar{0} \bar{0} \ldots \rangle -  | \bar{1} \bar{1} \ldots \rangle ) \; ,
\end{eqnarray*}
where the first one is a linear combination of standard basis
vectors with an even number of qubits up and the second one is  a
linear combination of standard basis vectors with an odd number of
qubits up
\begin{eqnarray*}
|g_1 \rangle = K_N \sum_{\{i,j, \ldots\}_e} | \{ i,j,\ldots \} \rangle \; ,\\
|g_2 \rangle = K_N\sum_{\{i,j, \ldots\}o} | \{ i,j,\ldots \} \rangle \; ,
\end{eqnarray*}
and the constant $K_N$ is a normalisation constant.
It is easy to show that these are the only two possible linear
combinations that satisfy the condition, and on top of that it is
clear from the construction that each of the states $|g_1 \rangle$
and  $|g_2 \rangle$ is a GHZ state.

\section{The  X-state}
\label{pstatea}

 The expression for  the X-state
 reads
\begin{equation}
\label{padef-p}
|X \rangle = \sum_{\{i,j,\ldots\}_e} | \{ i,j,k,\ldots \} \rangle (-1)^{ \sum_{i>j} d(i,j)} \; ,
\end{equation}
where $\{i,j, \ldots\}_e$ is a set of indices with an even number
of indices in the set so that we sum over all sets of indices with
an even number of indices, the letters $i,j,\ldots$ denote
positions of the qubits in the chain that are up, $d(i,j)$ is the
distance of the qubits on the $i$-th and $j$-th  positions defined
below and the sum over $i<j$ means that we sum over all pairs of
qubits up counting only once the cases with switched positions of
the qubits. Let us note that the state in Eq.~(\ref{padef-p})
is not normalised to unity.

\begin{figure}
\includegraphics[width=7cm]{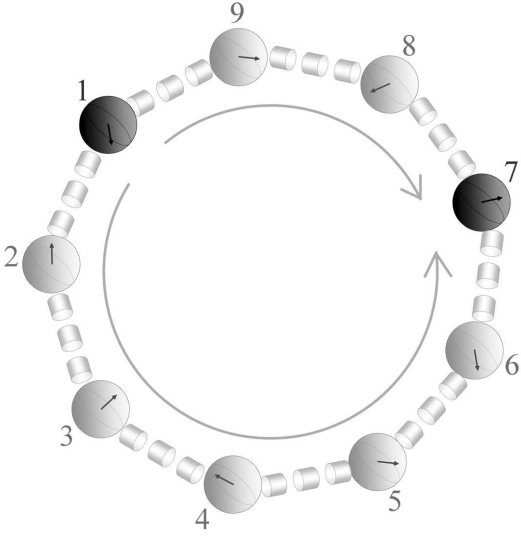}
\caption{The ring of 9 qubits. The arrows denote two possible paths from the
first qubit to the seventh qubit. The shorter path
is the distance between the two qubits $d(1,7)$. }
\label{chain9-1}
\end{figure}

The distance $d(i,j)$ of the two qubits is defined as the shortest
path on the ring  that brings us from the qubit on the $i$-th
positions to the qubit on the $j$-th position. As the qubits form
a circle, there are always two paths we can go without going
backward and we can always choose the shortest one. To make clear
what the distance defined above is, let us have a look at a simple
example. Let $N$ be $9$ so that overall number of qubits in the
ring is nine as in Fig.~\ref{chain9-1}.
 Further let $i=1$ so that
it denotes the first position and $j=7$ so that it denotes the
seventh position. Then the shortest path is going from first to
ninth position as the two are neighboring positions then from the
ninth to the eighth and finally from the eighth to the seventh
position. Consequently, the distance $ d(1,7)$ in this particular
case is $3$.

\subsection{Proof that the X-state is an eigenstate of the Hamiltonian with zero energy}

In what follows we show that the state is the eigenstate of the Hamiltonian
with zero energy or equivalently, that the following relation holds
\[  H |X \rangle = 0 \; .\]
If we divide the Hamiltonian into the free Hamiltonian and the interaction Hamiltonian $H= H_I+H_0$, then the last equation can be rewritten as
\begin{equation}
\label{rov}
 H_I |X \rangle = - H_0 |X \rangle \; .
\end{equation}
The task now is to show that the two vectors - one on the left and the other
on the right side of the last equation are equal. As we know, the equality
of two vectors follows from the equality of their components in any complete
basis. Actually what we will prove is the equality of the components of the
two vectors in the standard basis (computational basis) i.e.
\begin{equation}
\label{rov-komp}
 (H_I |X \rangle)_i = - (H_0 |X \rangle)_i \; .
\end{equation}
Take one vector of the standard basis that is included in the sum given by
Eq.~(\ref{padef-p}) and denote it as $|v \rangle$. We show that the $v$-th
components obey Eq.~(\ref{rov-komp}) \footnote{We know that in order to
prove  Eq.(\ref{rov}) we need to show that Eq.~(\ref{rov-komp}) holds for
all components. But the sum in Eq.~(\ref{def-p}) goes over all standard
basis vectors with an even number of qubits up. Moreover, as we will see
later neither $H_0$ nor $H_I$ can produce a vector with at least one nonzero
component of the standard basis vector with odd number of qubits up if the
standard basis decomposition of the vector we acted on does not contain a
vector with an odd number of qubits up. It follows that it is sufficient to
consider only components corresponding to the standard basis vectors with
even number of qubits up.} \mbox{} \\

\paragraph{}All vectors of the standard basis are eigenvectors of  $H_0$. If we denote $K$ to be the  number of qubits up in the vector $|v\rangle$ then
\begin{eqnarray*}
H_0 |v\rangle = (2K-N) |v\rangle\, ,
\end{eqnarray*}
and the $v$-th component of $H_0| X \rangle$ is
\begin{equation}
(H_0 |X \rangle)_v = ( 2K - N). s \, ,
\end{equation}
where $s$ is the phase factor of the vector $|v \rangle$ in the sum in
Eq.~(\ref{padef-p}). \mbox{} \\

\paragraph{}Now it remains to show that the same holds for $H_I$ except for the sign that must be opposite.
The Hamiltonian $H_I$ is a sum of many elements ${\bf C}_i$ where
\[ {\bf C}_i =  \sigma^x_i \sigma^x_{i+1} \; . \]
If we want to count the $v$-th component of $H_I|X \rangle$ we need to know
the individual contributions from each term ${\bf C}_i | X \rangle $. What
is the action of the operator ${\bf C}_i$? It flips two neighboring spins
on the $i$-th and $(i+1)$-th positions. Let us assume that there are $K$
spins up in the vector. If the two spins on the $i$-th and $(i+1)$-th
positions are up then the operator ${\bf C}_i$ flips them down and there are
$K-2$ spins up in the vector ${\bf C}_i |v \rangle $. Similarly if the two
spins on the $i$-th and $(i+1)$-th positions are down then the operator
${\bf C}_i$ flips them so that they are up and consequently, there are $K+2$
spins up in the vector ${\bf C}_i |v \rangle $. Otherwise if one spin is up
and the other is down then the number of spins up in the vector ${\bf C}_i
|v \rangle $ equals $K$.  At this point it is obvious that neither $H_0$ nor
$H_I$ can produce a contribution (vector of the standard basis) with an odd
number of qubits up since in Eq.~(\ref{padef-p}) we sum over all sets of
indices with an even number elements in each set. Now we use a little trick,
namely
\[ {\bf C}^2_i = (\sigma^x_i  \sigma^x_{i+1})^2 = {\bf 1}\, , \]
so that
\[ {\bf C}_i = {\bf C}_i^{-1} \; ,\]
in order to answer the question which vectors from the sum in
Eq.~(\ref{padef-p}) give contributions to the $v$-th element considering
only one ${\bf C}_i$. Using the last relation the only possible one is
\begin{equation}
\label{contribci}
 - s_i {\bf C}_i |v \rangle \; ,
\end{equation}
where $s_i$ is the coefficient of the state $ - {\bf C}_i |v \rangle$ in the
sum in Eq.~(\ref{padef-p}) (We have introduced the minus sign in the last
equation because of the sign of the operator ${\bf C}_i$ in $H_I$.). Since
the coefficients of the vectors in the sum Eq.~(\ref{padef-p}) are plus or
minus one there are exactly $N$ contributions to the $v$-th element, as we
have $N$ operators ${\bf C}_i$, and all of them are plus or minus one. Our
task is to find out the sign of each individual contribution and count them.
First we divide the contributions into three subsets. Let us
denote by $M_0$
the set of all vectors ${\bf C}_i |v \rangle$ with two more qubits up
compared with the vector $|v \rangle$. If we by denote $K_0$  the number
of pairs of neighboring qubits both being down in the vector $|v \rangle$
then the number of elements in the set $M_0$ is $K_0$. Equivalently,
let us denote
by $M_2$ to be the subset of all vectors ${\bf C}_i |v \rangle$ with two
more spins down as are in the vector $|v \rangle$. The number of elements in
the set $M_2$ is $K_2$ where $K_2$ is the number of pairs of neighboring
qubits both being up in the vector $| v \rangle$ . Finally,
let us  denote by $M_1$
the subset of all vectors ${\bf C}_i | v \rangle$ with equal number of
qubits up as are in the vector $| v \rangle$. The number of elements set
$M_1$ is $K_1$ and equals the number of pairs of neighboring qubits in the
state $|v \rangle $ with exactly one qubit up. The following relations hold
\begin{eqnarray}
\label{K+K+K=N}
K_1 + K_2 + K_3  & = & N\, ; \\
0. K_0 +1. K_1 + 2 . K_2 & = & 2 K\, . \nonumber
\end{eqnarray}
We can rewrite $s_i$ in the following way
$s_i = s . k_i$
where $k_i$ is the relative sign of the vector $ {\bf C}_i |v \rangle$
according to the absolute sign of the vector $| v \rangle$ in the sum
Eq.~(\ref{padef-p}).

Consider vectors belonging to the subset $M_0$.
To find out the relative sign $s_i$ between the vector  $ {\bf C}_i |v \rangle$  and vector $|v \rangle$ we need to know the following distances
\begin{enumerate}
\item
The distance from any qubit up in the vector $|v \rangle$ to the $j$-th
position: $d(j,x)$ .

\item
The distance from the $(j+1)$-th qubit to any qubit up in the vector $|v
\rangle$: $d(j+1,x)$ .

\item
The distance of the $j$-th and the $(j+1)$-th qubit that is apparently one:
$d(j,j+1) = 1$ .
\end{enumerate}
The relative sign between the two vectors is then
\begin{equation}
\label{suma}
(-1)^{1+ \sum_x d(j,x) + d(j+1,x)}\, ,
\end{equation}
where summing over x means that we sum over all positions of qubits up in
the vector $|v\rangle$. \\ \mbox{}

{\bf Theorem~2} {\em
If the qubit in the vector $|v \rangle$ that is equally distant from the
$j$-th and the $(j+1)$-th qubit is down then the exponent in
Eq.~(\ref{suma}) is odd.}

\mbox{} \\
{\em Proof:} There is only one qubit in the ring that is equally distant from the qubits on the $j$-th and the $(j+1)$-th positions. If that qubit is down and taking any qubit in the vector $|v \rangle$ that is up and summing the distance from the $j$-th qubit to the given qubit  and the distance from the $(j+1)$-th qubit to the same qubit we always get an odd number. As vector $|v \rangle$ contains even number of qubits up the sum is an even but to get the final expression in Eq.~(\ref{suma}) we have to add $1$ therefore the exponent is odd. \\ \mbox{}

{\bf Theorem~3} {\em
If the qubit that is equally distant from the $j$-th and the $(j+1)$-th qubits is up in the vector $|v \rangle$ then the exponent in the sum in Eq.~(\ref{suma}) is even. }

\mbox{} \\
{\em Proof:} Follows from the previous statement.

\mbox{} \\
We know that we have $K_0$ vectors in the set $M_0$. Denote $k_0$ to be the number of such states that they have the qubit which is equally distant from the corresponding $j$-th and $(j+1)$-th position up. Then we may say that the contribution of all vectors from the set $M_0$ to the $v$-th component is
\begin{equation}
\label{m0}
- (k_0 - (K_0 - k_0)) .s \; ,
\end{equation}
where the minus sign in front comes from Eq.~(\ref{contribci}).
It is amazing that considering the other sets namely, $M_1$ and $M_2$, we have come to the same conclusion so that their contributions to the $v$-th component are
\begin{equation}
\label{m1}
-(k_1 - (K_1 -k_1)) .s \; ,
\end{equation}
from $M_1$ and
\begin{equation}
\label{m2}
-(k_2 - (K_2 -k_2)) .s \; ,
\end{equation}
from $M_2$. Consequently, the $v$-th component of the vector $H_I|p \rangle$ is a sum of Eq.~(\ref{m0}), Eq.~(\ref{m1}) and Eq.~(\ref{m2}) and reads
\begin{eqnarray*}
( K_0 + K_1 +K_2 - 2 (k_0 +k_1 +k_2)) . s \: .
\end{eqnarray*}
Now comes the crucial point. The following equation holds
\begin{eqnarray*}
k_0 + k_1 + k_2 = K
\end{eqnarray*}
and together with Eq.~(\ref{K+K+K=N}) the $v$-th component of $H_I |p \rangle$ is
\begin{equation}
(H_I|X\rangle)_v = ( N -  2 K ) . s
\end{equation}

\paragraph{}
We have proved that the left-hand side of
Eq.~(\ref{rov-komp}) is equal to $ - ( N -  2 K ) . s$ and the
right-hand side of the equation equals  $ - ( N -  2 K ) . s$. In other
words the two expressions are equal for a given vector $|v
\rangle$. Since we have not specified the vector $|v \rangle$, it
holds for any vector (see the discussion above) and we have proved
that the X-state is an eigenstate of the Hamiltonian $H$ with zero
energy.

\subsection{Density matrix of n neighboring qubits}

{\bf Theorem~4} { \em
\label{pstaterho}
Let $N$ denote the total number of spins forming the  chain and $n$ be an integer. Further let  $N=2n+1$ and the system be in the X-state Eq.~(\ref{padef-p}).

Then the density matrix of  any sequence of $n$ neighboring qubits is
\begin{equation}
\rho_O = \frac{1}{2^n} {\bf 1}_n \; ,
\end{equation}
where ${\bf 1}_n $ is the identity operator acting on the $2^n$-dimensional
Hilbert space of $n$ qubits.
} \\ \mbox{}

The sequence of $n$ neighboring qubits is a subset of all qubits in the
chain such that only two ``cuts''
are needed to cut out the whole sequence from
the chain. In what follows we will denote the set of $n$ neighboring qubits
by $O$.
\vskip 0.3cm

\noindent {\em Proof of the Theorem~4:} We want to
show that the density operator of $n$ neighboring qubits is
proportional to the identity operator acting on the
$2^n$-dimensional Hilbert space. Denote by $|\{i, \ldots k \}
\rangle_O$ one of the basis state vectors of the system of $n$
qubits from the set $O$ where the set of indices $\{i, \ldots \}$
denote positions where the spins are up and all remaining spins
are down.
 First we rewrite the X-state using this new basis as follows
\begin{eqnarray}
\label{C13}
|X \rangle = \sum_{\{i, \ldots k \}} |\{i, \ldots j \} \rangle_O | \alpha_{\{i, \ldots j\}}
\rangle_{\bar{O}}\, ,
\end{eqnarray}
where $| \alpha_{\{i, \ldots j\}} \rangle_{\bar{O}}$ is a state vector of the remaining $n+1$ qubits
not belonging to the set $O$~\footnote{There are $N=2n+1$ qubits in the chain and only $n$ of them belong
to the set $O$.} and we sum over all sets of indices $\{i, \ldots k\}$
which means that we sum over all basis vectors of the system of $n$ qubits. Then the Theorem~4 says that
\begin{equation}
\label{otncon}
{}_{\bar{O}} \langle \alpha_{\{k, \ldots l\}} |\alpha_{\{i,\ldots j\}} \rangle_{\bar{O}} = \frac{K^2}{2^n} \delta_{\{k, \ldots l\}, \{i, \ldots j\}}\, ,
\end{equation}
where $K$ is the norm of the X-state.

In order to prove Eq.~(\ref{otncon}) we need to know the form of the states
$| \alpha_{\{i, \ldots j\}} \rangle_{\bar{O}}$. From Eq.~(\ref{padef-p}) we have
\begin{eqnarray*}
|X \rangle & = & \sum_{\{i,j,\ldots \}_e} | \{ i,j,k,\ldots \} \rangle (-1)^{ \sum_{i>j} d(i,j)} \\
& = &  \sum_{ \{ i, \ldots j \}_e } |\{i, \ldots j \} \rangle_O \sum_{ \{ k, \ldots l \} e} | \{k, \ldots l \} \rangle_{\bar{O}} \\
& & (-1)^{d_{\{i, \ldots j\}} + d_{\{k, \ldots l\}} + d_{\{i, \ldots j\}, \{ k, \ldots l \}} } \\
& + & \sum_{ \{ i, \ldots j \}o } |\{i, \ldots j \} \rangle_O \sum_{ \{ k, \ldots l \}o } | \{k, \ldots l \} \rangle_{\bar{O}} \\
& & (-1)^{d_{\{i, \ldots j\}} + d_{\{k, \ldots l\}} + d_{\{i, \ldots j\}, \{ k, \ldots l \}} } \; ,
\end{eqnarray*}
where the sum over  $\{i, \ldots j\}_e$
means that we sum over all sets of indices with an even number of indices in each set,
$\{j, \ldots k\}o$ means that we sum over all sets of indices with an odd number of indices in
each set, $d_{\{i, \ldots j\}}= \sum_{i>j} d(i,j) $ where $i,j \in \{i,j,\ldots \}$
and $d_{\{i, \ldots j\}, \{k, \ldots l\}} = \sum_{a \in \{i, \ldots j\}, b \in \{k, \ldots l\}} d(a,b)$.
It follows that the states  $| \alpha_{\{i, \ldots j\}} \rangle_{\bar{O}}$ are of the following form.
If the set $\{i, \ldots j \}$ contains an even number of indices then
\begin{eqnarray}
\label{alphaeven}
 |\alpha_{\{i,\ldots j\}} \rangle_{\bar{O}}
& = & \sum_{ \{ k, \ldots l \} e} | \{k, \ldots l \} \rangle_{\bar{O}}
\\ \nonumber  &\times &
(-1)^{d_{\{i, \ldots j\}} + d_{\{k, \ldots l\}} + d_{\{i, \ldots j\}, \{ k, \ldots l \}} } \; ,
\end{eqnarray}
while if the set contains an odd number of indices then
\begin{eqnarray}
\label{alphaodd}
 |\alpha_{\{i,\ldots j\}} \rangle_{\bar{O}} & = &
\sum_{ \{ k, \ldots l \} o} | \{k, \ldots l \} \rangle_{\bar{O}} \\ \nonumber
&\times & (-1)^{d_{\{i, \ldots j\}} + d_{\{k, \ldots l\}} + d_{\{i, \ldots j\}, \{ k, \ldots l \}} } \; .
\end{eqnarray}
Moreover, the norm of the X-state $K$ can be easily calculated and
the result is
\begin{eqnarray*}
K^2 & = & \langle X | X \rangle = \sum_{\{ i, \ldots j \}_e }  \sum_{\{ k, \ldots l \}_e } 1 + \sum_{\{ i, \ldots j \}o }  \sum_{\{ k, \ldots l \}o }  1 \\
& = & 2 2^{n-1} 2^n = 2^{2n} \; .
\end{eqnarray*}
In what follows we consider three different possibilities.

\begin{enumerate}
\item The two sets $\{i, \ldots j\}$,  $\{k, \ldots l\}$ are equal. If in the set $\{i, \ldots j\}$ is an even number of indices then
\begin{eqnarray*}
{}_{\bar{O}} \langle \alpha_{\{i, \ldots j\}} |  \alpha_{\{i, \ldots j\}} \rangle_{\bar{O}} = \sum_{\{ k, \ldots l \}_e } 1 = 2^n \; ,
\end{eqnarray*}
else if in the set  $\{i, \ldots j\}$ is an odd number of indices then
\begin{eqnarray*}
{}_{\bar{O}} \langle \alpha_{\{i, \ldots j\}} |  \alpha_{\{i, \ldots j\}} \rangle_{\bar{O}} = \sum_{\{ k, \ldots l \}o } 1 = 2^n \; .
\end{eqnarray*}

\item There is an even number of indices in the set  $\{i, \ldots j\}$ while in the set $\{k, \ldots l\}$ the number of indices is odd. If the set  $\{i, \ldots j\}$ contains an even number of indices from Eq.(\ref{alphaeven}) it follows that state  $ | \alpha_{\{i, \ldots j\}} \rangle_{\bar{O}}$ is a sum of basis state vectors with an even number of spins up. Further if  the set $\{k, \ldots l\}$ contains an odd number of indices then  from Eq.~(\ref{alphaodd}) it  follows that the state $ | \alpha_{\{k, \ldots l\}} \rangle_{\bar{O}}$ is a sum of state vectors with an odd number of spins up and therefore the scalar product Eq.~(\ref{otncon}) is zero i.e.
\[ {}_{\bar{O}} \langle \alpha_{\{i, \ldots j\}} |  \alpha_{\{k, \ldots l\}} \rangle_{\bar{O}} = 0 \; .\]
\item The case that remains is when in the both sets $\{i, \ldots j\}$ and $\{k, \ldots l\}$ there is an even number of indices but the two sets are
different~\footnote{The case with an odd number of indices in both sets but when the two
sets are different is equivalent to this case.}. The scalar product Eq.~(\ref{otncon}) reads as
\begin{eqnarray*}
& & {}_{\bar{O}} \langle \alpha_{\{i, \ldots j\}} |  \alpha_{\{k, \ldots l\}} \rangle_{\bar{O}}  = \hspace{1cm} \\
& & \sum_{ \{m, \ldots n\}_e} (-1)^{d_{\{i, \ldots j\}} + d_{\{m, \ldots n\}} + d_{\{i, \ldots j\}, \{ m, \ldots n \}} } \\
& & \hspace{1cm} \times (-1)^{d_{\{k, \ldots l\}} + d_{\{m, \ldots n\}} + d_{\{k, \ldots l\}, \{ m, \ldots n \}} } \\
& & =  (-1)^{d_{\{i, \ldots j\}}+ d_{\{k, \ldots l\}} }  \\
& & \times   \sum_{ \{m, \ldots n\}_e}  (-1)^{2  d_{\{m, \ldots n\}}+  d_{\{i, \ldots j\}, \{ m, \ldots n \}}+  d_{\{k, \ldots l\}, \{ m, \ldots n \}} } \; .
\end{eqnarray*}
Since we want to show that the scalar product of the two vectors is zero we drop the overall factor in front of the sum and using the relation $(-1)^{2 k} = 1$ we rewrite the last equation as follows
\begin{eqnarray*}
{}_{\bar{O}} \langle \alpha_{\{i, \ldots j\}} |  \alpha_{\{k, \ldots l\}} \rangle_{\bar{O}}  =  \hspace{1cm} \\
\sum_{\{m, \ldots n\}_e} (-1)^{ d_{\{i, \ldots j\}, \{ m, \ldots n \}}+  d_{\{k, \ldots l\}, \{ m, \ldots n \}} } \; .
\end{eqnarray*}
The problem of calculating the scalar product of two vectors has transformed into calculating the distances between the
two sets. If an index $a$ is in the both  sets of indices $\{i, \ldots j\}$ and $\{k, \ldots l\}$
then we can neglect it since we sum over distances from $a$ to $\{m, \ldots n \}$ twice and the term  $(-1)^{ 2 d_{\{a\}, \{m , \ldots n\}}} = 1$ does not  change the sign of the corresponding contributions. Therefore instead of calculating distances $ d_{\{i, \ldots j\}, \{ m, \ldots n \}}$  and  $d_{\{k, \ldots l\}, \{ m, \ldots n \}} $ , we create one set of indices
\begin{eqnarray*}
 D(\{i,\ldots j\}, \{k, \ldots l\}) & = & \{i,\ldots j\} \cup \{k, \ldots l\} \\
& - & \{i,\ldots j\} \cap \{k, \ldots l\} \; ,
\end{eqnarray*}
and then
\begin{eqnarray}
\label{prod}
{}_{\bar{O}} \langle \alpha_{\{i, \ldots j\}} |  \alpha_{\{k, \ldots l\}} \rangle_{\bar{O}}  = \hspace{1cm} \nonumber \\
\sum_{\{m, \ldots n\}_e} (-1)^{ d_{D(\{i,\ldots j\}, \{k, \ldots l\}), \{m, \ldots n\}}} \; .
\end{eqnarray}

It  is important to note that the set   $D(\{i,\ldots j\}, \{k, \ldots
l\})$ always contains only even number of indices~\footnote{Let us note that
this is also true in the case when there is an odd number of indices in the
two sets $\{i, \ldots j \}$ and $\{k, \ldots l\}$. Therefore from this point
the proof is thoroughly identical in both cases.}. The last step is to
calculate the distances.
In order to calculate these distances we use the following strategy:
We choose one of the positions from
$D(\{i,\ldots j\}, \{k, \ldots l\})$ such that it is the closest to the
qubits not belonging to $O$ and denote the position to be $a$. There are
only two qubits that are equally distant from $a$ and none of them belong to
$O$. Denote their positions as $b$ and $c$, where the position denoted as
$b$ is closer to the set $O$. Let us calculate the distances from
$D(\{i,\ldots j\}, \{k, \ldots l\})$ to $b$ and $c$. If the distance
is
\begin{eqnarray*}
d_{D(\{i,\ldots j\}, \{k, \ldots l\}), \{b\}} = C \, ,
\end{eqnarray*}
then
\begin{eqnarray}
\label{pacojaviem}
d_{D(\{i,\ldots j\}, \{k, \ldots l\}), \{c\}} = C+ (L-1)\, ,
\end{eqnarray}
where $L$ is the number of positions in $D(\{i,\ldots j\}, \{k, \ldots l\})$ as
\begin{eqnarray*}
d(i,c) &=& d(i,b) +1  \, ; \qquad
\forall i \in D(\{i,\ldots j\} \{k, \ldots l\}) /\ a \\
d(i,c) &=& d(i,c) \, ; \qquad i=a \; .
\end{eqnarray*}
{\bf Example:}
To make this clear let us consider a simple example $N=9$, $n=4$,
$O=\{1,2,3,4\}$ and $D(\{i,\ldots j\}, \{k, \ldots l\})=\{2,4\}$ see
Fig.~\ref{chain9-2}.
\begin{figure}
\includegraphics[width=7cm]{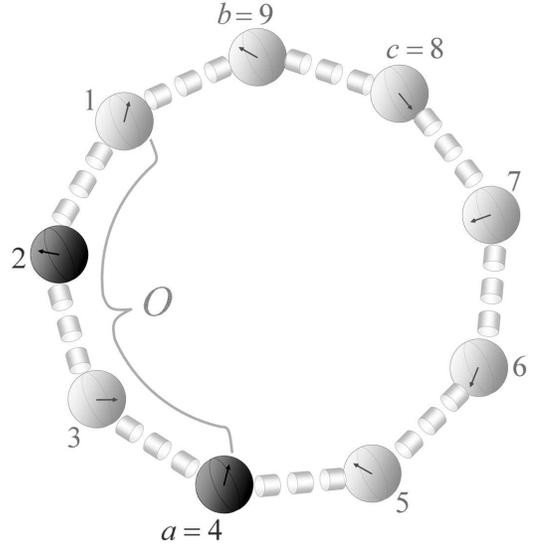}
\caption{The ring of 9 qubits. $O=\{1,2,3,4\}$.}
\label{chain9-2}
\end{figure}
The position that is the closest to the qubits not
belonging to $O$ is $4$ so that $a=4$. The two positions that are equally
distant from $4$ are $9$ and $8$. The position $9$ is closer to $O$ so that
$b=9$ and $c=8$. Moreover the distances
$d_{D(\{i,\ldots j\}, \{k, \ldots l\}), \{b\}}$ and $d_{D(\{i,\ldots j\},
\{k, \ldots l\}), \{c\}}$ are
\begin{eqnarray*}
d_{D(\{i,\ldots j\}, \{k, \ldots l\}), \{b\}} = d(2,9)+d(4,9) = 1 + 4 \; ,\\
d_{D(\{i,\ldots j\}, \{k, \ldots l\}), \{c\}} = d(2,8)+d(4,8) = 2 + 4 \; .\\
\end{eqnarray*}
The number of positions (indices) in $D(\{i,\ldots j\}, \{k, \ldots l\})$ is equal to $2$ and consequently the relation Eq.(\ref{pacojaviem}) holds.

Now we have everything necessary to calculate the scalar product Eq.~(\ref{prod}). Choose one arbitrary set  $\{m, \ldots n \}$ which contains neither $b$ nor $c$. If there is an even number elements in the set $\{ m, \ldots n \}$ then there are two contributions to
the sum in Eq.~(\ref{prod}) namely
\begin{eqnarray*}
(-1)^{ d_{D(\{i,\ldots j\}, \{k, \ldots l\}), \{m, \ldots n\}}} \\
+ (-1)^{ d_{D(\{i,\ldots j\}, \{k, \ldots l\}), \{b,c, m, \ldots n\}}} \; ,
\end{eqnarray*}
and these two have opposite signs [see Eq.~(\ref{pacojaviem})]. It follows
that these two contributions cancel each other.
Equivalently, if there is an odd number
of elements in the set $\{m, \ldots n \}$ then we have again two
contributions
\begin{eqnarray*}
(-1)^{ d_{D(\{i,\ldots j\}, \{k, \ldots l\}), \{a, m, \ldots n\}}} \\
+ (-1)^{ d_{D(\{i,\ldots j\}, \{k, \ldots l\}), \{b, m, \ldots n\}}} \; ,
\end{eqnarray*}
with opposite signs and the two contributions are mutually cancelled.
It follows that all contributions
to the sum in Eq.~(\ref{prod}) mutually cancel and the result is zero.
\end{enumerate}
To conclude,  we have proved that
\begin{eqnarray*}
{}_{\bar{O}} \langle \alpha_{\{k, \ldots l\}} |\alpha_{\{i,\ldots j\}} \rangle_{\bar{O}}  = 2^n \delta_{\{k, \ldots l\}, \{i, \ldots j\}} \; ,
\end{eqnarray*}
and it follows that the density operator of the $n$ neighboring qubits is
the identity operator up to a constant factor.

\subsection{Consequences}

\paragraph{Density matrix of a qubit}
Choose one  of $N$ qubits forming the Ising chain. Add  $n-1$ qubits to
the chosen qubit such that the $n$ qubits form a set of $n$ neighboring
qubits. Then according to Theorem~4 the density matrix of such
system is
\begin{eqnarray*}
\rho_n =  \frac{1}{2^n} {\bf 1}_n\, .
\end{eqnarray*}
When we now trace over the $n-1$ qubits that we added to the chosen qubit whose density matrix (state) we want to know, we obtain
\begin{equation}
\rho = {\rm Tr}_{n-1} \rho_n = \frac{1}{2} {\bf 1} \; ,
\end{equation}
where ${\bf 1}$ is the identity operator acting in the two-dimensional Hilbert space.

\paragraph{Density matrix of a pair of qubits}

The derivation of the density operator of any two qubits follows the same
steps as the derivation of the density operator of a single qubit. However,
in this case we add only $n-2$ qubits so that the set of $n$ qubits consists
only of $n$ neighboring qubits~\footnote{Let us note that for any pair
qubits we are able to choose other $n-2$ qubits so that the set of $n$
qubits is a set of $n$ neighboring qubits.}. According to
Theorem~4 the density operator of such system is $\rho_n =
1/2^n {\bf 1}_n$. The corresponding density operator of the two qubits is
obtained via tracing over the degrees of freedom of the $n-2$ qubits that we
added
\begin{equation}
\rho_{i,j} = Tr_{n-2}  \rho_n = \frac{1}{4} \; {\bf 1}_2 \; ,
\end{equation}
where $i,$ and $j$ denotes the positions of the two {\em a priori chosen} qubits
and ${\bf 1}_2$ is the identity operator acting in the four-dimensional
Hilbert space of the two qubits.

\end{document}